\documentclass[fleqn,usenatbib]{mnras}

\usepackage[T1]{fontenc}

\DeclareRobustCommand{\VAN}[3]{#2}
\let\VANthebibliography\thebibliography
\def\thebibliography{\DeclareRobustCommand{\VAN}[3]{##3}\VANthebibliography}

\usepackage{graphicx}	
\usepackage{amsmath}	
\usepackage{amssymb}	
\usepackage{wasysym}
\usepackage{caption}
\usepackage{subcaption}
\usepackage{hyperref}
\usepackage{iondefs}
\usepackage{threeparttable}
\usepackage{newtxtext,newtxmath}
\usepackage[]{xcolor}
\usepackage{deluxetable}
\usepackage{longtable}
\usepackage[font=small]{caption}
\newcommand{\MgIItitle}{\hbox{\bfseries Mg\kern0.1em\scshape ii}}
\newcommand{\MgIIhead}{\textit{Mg}\,\textsc{ii}}

\definecolor{my_red}{RGB}{255, 89, 94}
\definecolor{my_purple}{RGB}{143, 120, 174}
\definecolor{my_orange}{RGB}{255, 146, 76}
\definecolor{my_green}{RGB}{138, 201, 38}
\definecolor{my_blue}{RGB}{25, 130, 196}
\definecolor{my_yellow}{RGB}{255, 202, 58}
\definecolor{my_g1_green}{RGB}{100, 167, 71}
\definecolor{my_g1_blue}{RGB}{66, 103, 172}
\definecolor{my_pink}{RGB}{199, 105, 134}

\title[Tracing {\MgIIhead} gas flows]{Ultra-strong {\MgIItitle} absorbers trace both inflowing and outflowing gas: insights from dual down-the-barrel and quasar sightlines}

\author[A.~Fern\'{a}ndez-Figueroa et al.]
{Antonia Fern\'{a}ndez-Figueroa$^{1,2}$\thanks{E-mail: antoniafernandez@swin.edu.au},
Glenn G.~Kacprzak,$^{1,2}$ Tania M. Barone,$^{1,2}$ Nikole M.~Nielsen,$^{3,1,2}$
\newauthor{Kate H. R. Rubin$^{4}$,} Andrew J. Pitts$^{4}$, 
Barbara Mazzilli Ciraulo$^{1,2}$\\
$^{1}$Centre for Astrophysics and Supercomputing, Swinburne University of Technology, Hawthorn, Victoria 3122, Australia\\
$^{2}$ARC Centre of Excellence for All Sky Astrophysics in 3 Dimensions (ASTRO 3D), Australia\\
$^{3}$Homer L. Dodge Department of Physics and Astronomy, The University of Oklahoma, 440 W. Brooks St., Norman, OK 73019, USA\\
$^{4}$Department of Astronomy, San Diego State University, San Diego, CA 92182, USA
}

\date{Accepted 2025 October 2. Received 2025 October 1; in original form 2025 July 1}

\pubyear{2025}

\begin{document}
\label{firstpage}
\pagerange{\pageref{firstpage}--\pageref{lastpage}}
\maketitle

\begin{abstract}
We present Keck/LRIS spectroscopy of seven isolated galaxy--quasar pairs at $0.4 \leq z \leq 0.6$, each exhibiting ultra-strong {\MgII} absorption ($W_{r,2796} \geq 3$~\AA), probing both down-the-barrel and transverse gas flows. Down-the-barrel galaxy spectra reveal outflows in three galaxies ($v = 19$ to $311$ {\kms}) and inflows in five ($v = 61$ to $361$ {\kms}), including one system showing inflows and outflows simultaneously. All galaxies with detected inflows are below the star-forming main sequence, suggesting that they might be actively replenishing their gas reservoirs. Outflows have a mean covering fraction of $C_{f, \rm out}=0.5$, whereas inflows show a lower average of $C_{f, \rm in}=0.3$. Mass flow rates span $\dot{M}_{\rm in} = 0.01-1.18 M_{\odot} \mathrm{yr}^{-1}$ for inflows and $\dot{M}_{\rm out} = 0.23-1.03 M_{\odot}\mathrm{yr}^{-1}$ for outflows, yielding mass loading factors below unity and implying these galaxies cannot sustain their current level of star-formation rates. These results are based on the T~$\sim 10^4$~K photoionised gas phase traced by {\MgII}; additional accreting/outflowing material in other gas phases may also be present, but remains undetected in this study. Quasar sightlines consistently show redshifted inflow components and blueshifted outflow components, demonstrating that ultra-strong {\MgII} absorbers trace baryon cycling out to impact parameters of $D = 15$ -- 31~kpc. Moreover, the unexpectedly high prevalence of inflows suggests that ultra-strong {\MgII} absorbers offer a powerful strategy for future surveys to systematically map inflow and outflow cycles across cosmic time.
\end{abstract}




\section{Introduction}

The circumgalactic medium (CGM) is a giant reservoir of gas that contains $50$\% of the baryonic matter in the Universe \citep{Werk2014}. This giant gas reservoir hosts a set of processes associated with the baryon cycle, such as galactic scale outflows and accretion. Studying the CGM is fundamental for the understanding of galaxy evolution, however, its diffuse nature makes observational studies challenging. Although some recent studies have managed to map the CGM in emission \citep{Burchett2021, Zabl2021, Leclercq2022, Nielsen2024, Pessa2024}, this galaxy component is typically detected in absorption using bright quasars (QSOs) as background sources \citep[e.g.][]{Lanzetta_1995, Chen_1998a, Churchill_2000a, Kacprzak2008, Nielsen_2013b, Schroetter2016, Prochaska_2017b, Tumlinson_2017, Dutta2020, Huang2021, Peroux2022}. Additionally, several studies have used the galaxy's own light to investigate their CGM, a technique known as down-the-barrel spectroscopy \citep{Rubin2012, Rubin_2014, Martin2012, Kornei2012, Kacprzak2014, Heckman2015}.

The {\MgII} doublet has historically been used as a tracer for cool ($T\sim 10^{4}$~K) gas in the CGM \citep{Bergeron&Boisse1991, Steidel1994, Churchill_2000a, Churchill2005, Rigby2002, Chen_2010a, Kacprzak2011b}. This is partly due to the fact that it lies in the optical wavelengths at redshifts $0.1<z<2.5$ and because it traces gas associated with ${16\leq \log \mathrm{N}({\HI}) / {\rm cm}^{-2}\leq22}$ \citep{Bergeron&Stasinska1986, Steidel&Sargent1992, Churchill1999, Churchill_2000a, Rao&Turnshek2000, Rigby2002} at distances up to $150$~kpc from its host galaxy \citep{Kacprzak2008, Chen_2010a, Churchill2013a, Nielsen2013a}. Due to these properties, {\MgII} absorption lines have been used to study both inflows \citep{Kacprzak_2010a, Kacprzak2011b, Kacprzak2012, Ribaudo2011, Stewart2011, Martin2012, Rubin2012, Ho2017, Rahmani_2018, Zabl2019} and outflows \citep{Bouche2006, Bouche2012, Tremonti2007, Martin&Bouche2009, Weiner2009, Chelouche2010, Rubin2010, Bordoloi_2011a, Coil2011, Martin2012, Kacprzak2012, Zhu&Menard2013, Nielsen_2013b, Rubin_2014, Schroetter2019, Schroetter2024} in the CGM.

Down-the-barrel spectroscopy has proven to be an effective method for detecting blueshifted absorption lines in galaxies with respect to their systemic velocity, traditionally interpreted as signatures of outflows \citep{Rubin_2014}. In contrast to outflows, which typically exhibit velocities between $50$ and $800$ {\kms} \citep{Steidel2010, Rubin_2014, Du2016}, inflows have lower velocities \citep[$\sim$ $100$ to $200$ {\kms},][]{Goerdt_Ceverino2015, Carr_Scarlata2022}, making them more difficult to disentangle from ISM absorption, particularly in low resolution spectra. In this context, redshifted absorption features detected down-the-barrel are generally attributed to inflows or gas accreting through corotating disks. Early detections of inflows using this technique were primarily performed in galaxies hosting active galactic nuclei \citep[AGN, ][]{LeFloch2007, Krug2010, Stone2016}, early-type \citep{Sato2009, Sarzi2016} and post-starburst galaxies \citep{Coil2011}.

Although inflows are not often detected, understanding their origins has been a key goal of down-the-barrel spectroscopy studies. \citet{Rubin2012} presented the first detection of inflows using down-the-barrel spectroscopy on star-forming galaxies, identifying six cases within a sample of $\sim 100$ galaxies at $0.3 \leq z \leq 1.4$. An analysis of the morphologies of the galaxies associated with these inflows revealed that five out of six have high inclinations, supporting the idea that accretion tends to occur parallel to the galaxy disk \citep[e. g. ][]{Bouche2013}. \citet{Martin2012} performed a similar analysis in a sample of 208 star-forming galaxies at the same redshift range and detected inflows in nine of them. However, only one of these galaxies had a high inclination, and the authors attributed the inflows to minor mergers instead of corotating gas. More recent studies have also explored the possibility that these inflows are tracing gas that was originally ejected from their host galaxies, has cooled down and is falling back into the galaxy, a process known as ``galactic fountain'' \citep{Zheng2017, Roberts-Borsani2019, Bolda2024}.

Ultra-strong {\MgII} absorbers are defined as systems with {\MgII} rest-frame equivalent width higher than $3$~\AA~\citep{Nestor2007}. These systems are rare, comprising only about $0.8$\% of all systems found in quasar absorption line surveys \citep{Zhu&Menard2013, Guha2022}. They have extremely high velocity dispersions, often reaching several hundred {\kms}, exceeding the escape velocity of individual halos, and cannot be explained by large column densities or saturation in the absorption. It has been thoroughly shown in the literature that {\MgII} absorption decreases as impact parameter increases \citep{Nielsen_2013b, Kacprzak2013, Dutta2020, Huang2021}, but ultra-strong absorbers seem to defy that relation \citep{Guha2024, Joshi2024}, remaining out to large impact parameters ($D\sim 40$~kpc).

Several hypotheses have been proposed to explain the origin of ultra-strong {\MgII} absorbers. One possibility is that they arise from interactions between galaxies, which can strip cool gas to large distances from their hosts \citep{Gauthier2013, Nielsen2022}. This scenario is supported by the finding that nearly $70$\% of these absorbers are associated with galaxy groups \citep{Guha2022, Guha2024}, similar to the flattening of the {\MgII} equivalent width-impact parameter relation observed around groups \citep{Nielsen_2018b, Fossati2019, Dutta2021, Huang2021}. Another possibility is that they are produced by star formation-driven outflows \citep{Rubin2010, Nestor2011, Nielsen2022, Joshi2024}. In many systems, evidence for both processes is present, making it difficult to determine which mechanism predominantly drives the ultra-strong absorption. Notably, the existence of ultra-strong absorbers associated with isolated galaxies \citep{Guha2022, Guha2024} suggests that group interactions cannot be the sole explanation.

There is evidence to suggest that star formation-driven outflows may be the main mechanism producing ultra-strong {\MgII} absorption. Several studies have identified a correlation between {\MgII} equivalent width and galaxy {\OIIe} luminosity \citep{Zibetti2007, Noterdaeme2010, Menard2011}. Given that star formation rates (SFR) are proportional to {\OIIe} luminosity \citep{Kewley_2004d}, this implies a correlation between {\MgII} equivalent width and SFR, particularly for the strongest absorbers ($W_{r,2796} > 1$~\AA) as shown by \citet{Menard2011}. \citet{Rubin_2014} also supported this connection by finding a positive correlation between {\MgII} outflow equivalent width and SFR in down-the-barrel observations. This makes the link especially relevant for ultra-strong {\MgII} systems. In addition, galaxies with higher star formation rate surface densities ($\Sigma_{\mathrm{SFR}}$) are more likely to generate the energy required to launch strong outflows \citep{Heckman2002, Heckman2015, ReichardtChu2022}, further supporting this scenario. The observed relation between {\MgII} absorption strength and SFR provides indirect evidence that outflows play a significant role in producing these absorbers.

Recently, \citet{Guha2024} studied $18$ ultra-strong {\MgII} absorbers at $0.6 < z < 0.8$. The impact parameters of these absorbers turned out to be much larger than expected, 
defying the well-known anti-correlation between {\MgII} equivalent width and impact parameter. Additionally, they found that, for a given impact parameter, the host galaxies of ultra-strong {\MgII} absorbers are more massive and have higher rest-frame $B$-band absolute magnitudes. In their study, at least $29$\% of the host galaxies are isolated, indicating that the ultra-strong absorption would likely come from outflowing material. However, the SFR of these galaxies is slightly lower than main sequence galaxies with similar stellar masses, suggesting that these galaxies could be transitioning from being star-forming to being quiescent.

To further investigate the origin of ultra-strong {\MgII} absorbers, we selected a sample of seven absorbers from \citet{Guha2022}, all associated with isolated galaxies. The isolation criterion helps eliminate the added complexity of controlling for group environments. We performed Keck/LRIS long-slit observations of the galaxies and the QSOs associated with each absorber to search for signatures of inflows and outflows. This observational setup allows us to simultaneously observe absorption down-the-barrel of the galaxies and absorption along the quasar sightlines, offering complementary measurements of inflows and outflows in and around these galaxies.

This paper is organised as follows: Section~\ref{data_and_methods} details our methods for data acquisition, reduction and analysis. General properties of the sample are described in Section~\ref{gen_prop}. We present results on the down-the-barrel analysis in Section~\ref{dtb_results}, where we detect inflows and outflows, and we later compare those to the absorption observed in the quasar absorption lines in Section~\ref{quasar}. In Section~\ref{discussion} we discuss the possible causes of the observed ultra-strong {\MgII} absorption, the most likely origins of our detected inflows and explanations for the prevalence of inflows in our sample. Finally, Section~\ref{conclusions} contains a summary and conclusions of the paper. We assume a flat $\Lambda$ cold dark matter ($\Lambda$CDM) Universe with $\Omega_{\Lambda}= 0.7$, $\Omega_{M} = 0.3$, and $H_{0} = 70$~km~s$^{-1}$~Mpc$^{-1}$, and a Kroupa initial mass function \citep{Kroupa_2001b}.

\section{Data and methods} \label{data_and_methods}

\subsection{Sample}

\begin{table*}
	\centering
	\caption{Properties of the foreground galaxies.}
	\label{tab:properties}
	\begin{tabular}{lcccccccccc} 
     \hline
        Quasar & RA & DEC & $z_{\rm gal}$ & $D$$^{a}$ & $W_{r,2796}$ & SFR & $\Sigma_{\mathrm{SFR}}$ & $\log(M_{\star}/M_{\astrosun})$$^{a}$ & $\log(M_{h}/M_{\astrosun})$$^{a}$ & $R_{\rm h}$$^{a}$\\
         &  &  &  & (kpc) & (\AA) & ($M_{\astrosun}$~yr$^{-1}$) & (M$_{\astrosun}$~yr$^{-1}$~kpc$^{-2}$) &  & & (kpc)\\
        (1) & (2) & (3) & (4) & (5) & (6) & (7) & (8) & (9) & (10) & (11) \\
	\hline
        J000413 & 00:04:13.99 & $-$08:26:24.27 & 0.55389 & 26 & 3.12$\pm$0.50 & 2.3$\pm$0.2 & 0.004 & $10.99_{-0.06}^{+0.06}$ & 12.24 & 228\\[2pt]
        J024008 & 02:40:08.39 & $-$08:12:22.53 & 0.53078 & 18 & 3.36$\pm$0.02 & 6.6$\pm$1.2 & 0.03 & $10.57_{-0.08}^{+0.07}$ & 11.87 & 174\\[2pt]
        J033438 & 03:34:38.08 & $-$07:11:52.08 & 0.59859 & 28 & 3.62$\pm$0.02 & 33.2$\pm$1.8 & 0.18 & $10.57_{-0.14}^{+0.12}$ & 11.88 & 169\\[2pt]
        J092222 & 09:22:22.52 & $+$04:08:56.00 & 0.45433 & 15 & 3.18$\pm$0.01 & 1.0$\pm$0.1 & 0.007 & $10.26_{-0.07}^{+0.09}$ & 11.66 & 154\\[2pt]
        J121453 & 12:14:53.46 & $+$08:04:57.03 & 0.49098 & 16 & 3.42$\pm$0.01 & 2.7$\pm$3.2 & 0.009 & $10.71_{-0.17}^{+0.13}$ & 11.97 & 192\\[2pt]
        J155003 & 15:50:03.47 & $+$03:13:25.77 & 0.56913 & 23 & 2.94$\pm$0.11 & 1.7$\pm$0.2 & 0.002 & $11.62_{-0.20}^{+0.19}$ & 13.56 & 624\\[2pt]
        J220330 & 22:03:29.76 & $-$00:22:15.03 & 0.43733 & 31 & 4.17$\pm$0.15 & 3.6$\pm$4.3 & 0.01 & $10.22_{-0.12}^{+0.11}$ & 11.64 & 153\\
        \hline
\end{tabular}

\raggedright The columns are: (1) Quasar name, (2) galaxy right ascension, (3) galaxy declination, (3) galaxy redshift, (5) impact parameter, (6) measured \MgII$\lambda2796$ rest-frame equivalent width of the QSO absorption, (7) extinction-corrected, \Hbeta~derived star-formation rate, (8) calculated $\Sigma_{\mathrm{SFR}}$, (9) galaxy stellar mass, (10) galaxy halo mass, (11) galaxy halo radius. \\ $^{a}$\footnotesize{Adopted from \citet{Guha2022}.}
\end{table*}

The sample consists of seven galaxies that reside within $15-31$~kpc of a background quasar and are associated with ultra-strong {\MgII} absorption, defined as systems with rest-frame equivalent width higher than 3 \AA. These seven galaxies were studied in \citet{Guha2022}, who identified them as isolated using Sloan Digital Sky Survey \citep[SDSS;][]{York2000} imaging. The isolation criterion required no other galaxies within $50$~kpc of the quasar line of sight down to a magnitude limit of $m_{r} < 23.6$. The galaxies are all star-forming with SFRs spanning from $1$ to $34$~M$_{\odot}$~yr$^{-1}$, stellar masses in the range of $\log M_{*}/M_{\odot} = 10.2-11.6$,
and redshifts between $0.4 < z < 0.6$. Galaxy properties are listed in Table \ref{tab:properties}.

\subsection{Keck/LRIS spectroscopy}

We obtained observations of both galaxy and quasar for each field on 15 November 2023 UT (PID: 2023B\_W431) and 3 March 2025 (PID: 2025A\_W368) using the Low-Resolution Imaging Spectrometer \citep[LRIS;][]{OKE1995}. We used the $600$/$4000$ grism on the blue CCD, along with the $900$/$5500$ grating on the red CCD, resulting in a wavelength range of $3115$ to $5650$ {\AA} on the blue CCD and $5475$ to $8150$ {\AA} on the red CCD. This range covers {\MgII}, {\MgI}, {\FeII}, and {\CaII} absorption, along with {\OIIIe} and {\Hbeta} emission for all of our targets. However, the detectability of {\OIIe} emission varies across our sample. In four of our galaxies (J000413, J024008, J092222 and J220330), {\OIIe} emission falls in a region of the blue side of the spectrum with low throughput due to the dichroic cutoff after 5000~\AA, where the instrument throughput significantly drops, making the lines undetectable. In contrast, for the other three (J033438, J121453 and J155003), {\OIIe} emission lay on the red side of the spectrum, allowing for a clear detection. The location of the LRIS slits is displayed in the top left panels of Fig.~\ref{fig:mgii_outflow}. The slit configuration was chosen to collect the quasar and galaxy simultaneously in each exposure. The exposure times averaged 2000~s on the blue CCD and 950~s in the red CCD. The observations were taken under seeing conditions of 0.7$''$ on both nights.

We reduced the Keck/LRIS data using the PypeIt software \citep{Prochaska2020} with its standard routines. After obtaining the reduced 2D spectra, we coadded all the exposures for each target. The coadded galaxy spectra achieved a mean SNR of 21 in the continuum around the {\MgIIdblt} absorption lines. We then extracted the 1D spectra of the quasar and galaxy on each of our seven selected fields manually. The extracted spectra were barycentre-corrected with vacuum wavelengths.

\subsubsection{Flux calibration}

We chose to perform flux calibrations by utilising the already flux-calibrated SDSS spectra of the quasar, since our LRIS spectra cover both the quasar and the targeted galaxy. This method is preferred over the typical standard star approach because the quasar is in the same slit as the galaxy throughout the observations.

Given that the quasar and galaxy have different morphologies (point source versus extended), we must consider light loss from the slit differently.  We accounted for slit-loss by using SDSS and Dark Energy Camera Legacy Survey \citep[DECaLS,][]{Dey2019} imaging, which closely match the seeing of our observations. We first scale the SDSS aperture-derived flux calibrated spectra down by the ratio of the image counts within the slit and the counts within the SDSS aperture. This yields a SDSS flux calibrated spectrum as observed through our LRIS slit. We then match the LRIS quasar spectra to the scaled SDSS spectra to flux calibrate them. This scaling is then applied to the galaxy, which provides the flux of the galaxy within the slit only. We then scale the galaxy spectra by the ratio of the image counts of the total galaxy over the galaxy counts within the slit. By doing so, this accounts for the different sizes and shapes of the galaxies. This second scaling provides the total flux of the entire galaxy. This flux calibration was tested on J220330, which is the only galaxy with an available SDSS spectrum. We find that the flux difference between our galaxy-calibrated spectrum and the SDSS spectrum was only $13$\%.

\subsubsection{Extinction corrections}

The Galactic extinction was corrected using the method described in \citet{Cardelli_1989}. The $A_{\rm V}$ values at the RA and DEC of the targets were found in the NED Extinction Calculator \footnote{\url{https://ned.ipac.caltech.edu/extinction_calculator}}.

The Balmer series lines in our galaxies displayed significant stellar absorption on top of the gas emission, so it was necessary to apply a continuum subtraction to recover their fluxes. The continuum was fitted using the pPXF software \citep{Capellari_2004, Capellari_2017a, Capellari_2022}, which is designed to model stellar population properties using the penalised pixel-fitting algorithm.  We used the stellar templates from the Extended Medium resolution INT Library of Empirical Spectra \citep[E-MILES,][]{Sanchez_Blazquez_2006b}, with BaSTI isochrones \citep{Pietrinferni2004, Pietrinferni2006}, and a Kroupa IMF. These templates span a range of $53$ ages from $0.03$ to $14.00$ Gyr, $12$ [Z/H] values from $-2.27$ to $0.40$, and have [$\alpha$/Fe] values scaled to the solar neighbourhood \citep[base models;][]{Vazdekis2016}. The software was run with an additive polynomial of order $12$ \citep[see][]{VanDeSande2017}. The continuum fitted by pPXF was then subtracted from the galaxy spectra. To ensure that the continuum levels were actually zero around the emission lines, we fitted another three-degree polynomial and subtracted this from the spectrum.

Lastly, we corrected for the intrinsic dust extinction of each galaxy using once again the method described in \citet{Cardelli_1989}. The $E(B-V)$ values were calculated by comparing our measured {\Hbeta} and {\Hgamma} flux ratios to the intrinsic ones, as explained in \citet{Calzetti_2001}. We then transformed these $E(B-V)$ values into $A_{\rm V}$ by assuming a total-to-selective extinction ratio $R_{\rm V} = 3.1$, as is standard practice. With these dust-corrected lines we are able to compute SFRs, metallicities and emission line ratios.

\subsection{Star-formation rates}

The star-formation rate was calculated using the {\Hbeta} luminosity with the following equation:
\begin{equation}
    {\rm SFR} = C_{\tiny \Halpha} \frac{L_{\tiny \Halpha}}{L_{\tiny \Hbeta}} 10^{-0.4 A_{\tiny \Hbeta}} L'_{\tiny \Hbeta},
\end{equation}
where ${L_{\tiny \Halpha}}/{L_{\tiny \Hbeta}}=2.87$ is the intrinsic luminosity ratio \citep{Calzetti_2001}, $L'_{\tiny \Hbeta}$ is the observed {\Hbeta} luminosity, and $C_{\tiny \Halpha}=10^{-41.257}$ is the scale parameter from \citet{Hao_2011b}, assuming a Kroupa IMF \citep{Kroupa_2001b}, $Z_{\astrosun}$, 100 Myr model. The extinction parameter $A_{\tiny \Hbeta}$ was assumed to be 0 since the dust extinction correction had already been performed. The computed SFRs are listed in Table~\ref{tab:properties}. 

\subsection{ISM metallicity and emission line ratios} \label{metallicity}

The interstellar medium (ISM) metallicity was calculated using the method from \citet{Kewley_2004d}, which estimates oxygen abundances based on the $R_{23}$ and $O_{32}$ parameters. Here {$R_{23}=(\OIIe\lambda3727 + \OIIIe\lambda\lambda4959,5007)/\Hbeta$}, and {$O_{32}=\OIIIe\lambda\lambda4959,5007/\OIIe\lambda3727$}. An initial guess of the metallicity is generated by averaging the values obtained using the \citet{Zaritsky_1994} and \citet{McGaugh_1991} methods, and this value is used to decide which branch of $R_{23}$ is appropriate to use. Unfortunately, only three of our targets (J033438, J121453 and J155003) have {\OIIe} detections in our LRIS data, so it was only possible to calculate metallicities for their associated galaxy. This metallicity value will later be used to calculate oxygen column densities and mass outflow rates.

To distinguish AGNs from star-forming galaxies, we used the method described in \citet{Lamareille2010}, which uses the  \OIIe/\Hbeta~and \OIIIe/\Hbeta line ratios. This analysis of J033438, J121453 and J155003 showed that they do not have AGN activity.

\subsection{Absorption line modelling} \label{vp_fitting}

\begin{table*}
	\centering
	\caption{Properties of the down-the-barrel flows.}
	\label{tab:flows}
	\begin{tabular}{lccccccc} 
      \hline
      Target & $W_{r,2796}$ & \ensuremath{\log (N(\MgII)/{\rm cm}^{-2})} & $\Delta$v & $b_{\rm D}$ & $C_{f}$ & $\dot{M}$ & $\eta$ \\
      & (\AA) &  & (\kms) & (\kms) &  & ($M_{\astrosun}$~yr$^{-1})$ & \\
      & (1) & (2) & (3) & (4) & (5) & (6) & (7)\\
      \hline
      & \multicolumn{6}{c}{Inflow}\\
      \hline
      J000413 & $0.77\pm0.11$  & $\geq 15.0$
      & $261_{-17}^{+15}$ & $88_{-31}^{+51}$ & $0.22_{-0.04}^{+0.09}$ & $\geq 0.60$ & $\geq 0.26$\\[2pt]
      J024008a$^{a}$ & $0.32\pm0.07$ & $\geq 13.6$
      & $61_{-40}^{+40}$ & $73_{-34}^{+40}$ & $0.37_{-0.17}^{+0.42}$ & $\geq 0.01$ & $\geq 0.001$\\[2pt]
      J024008b$^{a}$ & $0.26\pm0.10$ & $\geq 14.4$
      & $361_{-15}^{+16}$ & $75_{-34}^{+66}$ & $0.16_{-0.06}^{+0.43}$ & $\geq 0.08$ & $\geq 0.01$\\[2pt]
      J092222 & $0.82\pm0.06$  & $\geq 15.3$
      & $189_{-17}^{+17}$ & $69_{-24}^{+44}$ & $0.33_{-0.06}^{+0.09}$ & $\geq 0.66$ & $\geq 0.66$\\[2pt]
      J155003 & $1.84\pm0.23$ & $\geq 14.91$
      & $213_{-3}^{+4}$ & $199.6_{-0.7}^{+0.3}$ & $0.40_{-0.00}^{+0.01}$ & $\geq 0.30$ & $\geq 0.18$ \\[2pt]
      J220330 & $2.31\pm0.20$ & $\geq 14.6$
      & $299_{-19}^{+19}$ & $141_{-36}^{+23}$ & $0.61_{-0.10}^{+0.11}$ & $\geq 0.37$ & $\geq 0.10$ \\
      \hline
      & \multicolumn{6}{c}{Outflow}\\
      \hline
      J000413 & $0.54\pm0.02$ & $\geq 14.8$
      & $-273_{-19}^{+19}$ & $54_{-24}^{+119}$ & $0.29_{-0.19}^{+0.10}$ & $\geq 1.64$ & $\geq 0.71$\\[2pt]
      J033438 & $3.15\pm0.11$ & $\geq 14.4$
      & $-311_{-10}^{+9}$ & $189_{-10}^{+7}$ & $0.75_{-0.05}^{+0.06}$ & $\geq 0.05$ & $\geq 0.001$\\[2pt]
      J121453 & $0.50\pm0.40$ & $\geq 13.4$
      & $-119_{-61}^{+55}$ & $131_{-65}^{+46}$ & $0.43_{-0.31}^{+0.41}$ & $\geq 0.003$ & $\geq 0.001$\\
      \hline
\end{tabular}

\raggedright The columns are: (1) rest-frame \MgII$\lambda$2796 equivalent width of the flow component, integrated over the fitted model described in Section~\ref{vp_fitting}, (2) \MgII~column density of the flow, (3) velocity of the inflow/outflow with respect to the ISM of the galaxy, (4) Doppler parameter of the flow, (5) covering fraction, (6) mass inflow/outflow rate, (7) mass loading factor. \\ $^{a}$\footnotesize{J024008 features two inflow components. The bluer/redder component is listed as J024008a/J024008b.}
\end{table*}

\subsubsection{Background quasar absorption lines}

The {\MgII} absorption lines observed in the quasar spectra are modelled using pairs of Gaussian profiles, each representing the $\lambda2796$ and $\lambda2803$ lines, with the same central velocities and velocity dispersions. The central velocity of the Gaussian pairs are allowed to vary within $-500$ and $500$ {\kms} from the systemic velocity of its host galaxy. Additional Gaussian pairs are added iteratively until a satisfactory fit is reached, upon visual inspection. None of these pairs is constrained to be at the systemic velocity of the host galaxy. Rest-frame equivalent widths for the full absorption profiles are listed in Table~\ref{tab:properties}.

\subsubsection{Down-the-barrel absorption}

We model the down-the-barrel {\MgII} absorption as a combination of two components: one associated with the galaxy's ISM, and another blueshifted or redshifted component, associated with outflows or inflows, respectively. The ISM component is fixed to have the same central velocity and velocity dispersion as the {\Hbeta} emission line, while the other parameters are left free. After fitting, we subtract the ISM contribution to the {\MgII} absorption in order to analyse the flow component(s) in detail.

The \MgII~column densities, Doppler parameters and covering fractions of the leftover blueshifted or redshifted absorption were calculated using a model adopted from \citet{Rubin_2014}. The model describes the normalized flux line intensity as $I(\lambda) = 1 - C_f(\lambda) + C_f(\lambda)e^{-\tau (\lambda)}$, where $C_f(\lambda)$ is the gas covering fraction and $\tau(\lambda)$ is the optical depth. We assume that $C_f(\lambda)$ is constant for each model as we do not attempt to constrain the contribution of individual gas clouds along the line of sight. The optical depth is assumed to take the form of a Gaussian,
\begin{equation}
    \tau(\lambda)=\tau_0 \mathrm{exp}\left[ -\frac{(\lambda - \lambda_0)^2}{(\lambda_0 b_D/c)^2} \right],
\end{equation}
where $\lambda_0$ and $\tau_0$ are the central wavelength and central optical depth of the line, respectively, and $b_D$ is the Doppler parameter. We may write the central optical depth of a given transition in terms of the column density of the gas, $N$, via
\begin{equation}
    \tau_0 = \frac{f_0}{\left( 1.497\times10^{-15} \right) \lambda_{\mathrm{rest}} \left[ \mathrm{\Angstrom} \right]} \frac{N \left[ \mathrm{cm^{-2}}\right]}{b_D \left[ \mathrm{km\ s^{-1}} \right]},
\end{equation}
where $f_0$ is the oscillator strength of the transition. By expressing the total optical depth of the \MgIIdblt~transition as the sum of its two components, $\tau(\lambda) = \tau_{\mathrm{blue}}(\lambda) + \tau_{\mathrm{red}}(\lambda)$, both lines are modelled simultaneously. We link the components of the optical depth by using the known oscillator strength ratio, $\tau_{\mathrm{blue,0}} / \tau_{\mathrm{red,0}} \approx f_{\mathrm{blue,0}} / f_{\mathrm{red,0}} \approx 2$, and assume the same $b_D$ for both lines. We also fix the separation of the doublet by treating only the central wavelength of the redder line as a free parameter. The central wavelength of the bluer line is determined by a fixed velocity offset, derived from the rest-frame wavelengths of the two transitions. Specifically, we relate the two line centres using
\begin{equation}
    \lambda_{\mathrm{blue,0}} = \lambda_{\mathrm{red},0} \left( 1 + \frac{2796.35\ \Angstrom - 2803.53\ \Angstrom}{2803.53\ \Angstrom}\right)
\end{equation}
where $2796.352\ \Angstrom$ and $2803.53\ \Angstrom$ are the vacuum rest-frame wavelengths of the \MgII~doublet \citep{Cashman2017}. Thus, our model has the four free parameters $C_f$, $b_D$, $N$, and $\lambda_{\mathrm{red,0}}$. To simulate the spectral resolution of the observations, we convolve the synthetic flux with a Gaussian kernel of $\mathrm{FWHM = 4}$~\AA, corresponding to the resolution element of Keck/LRIS. The smoothed spectrum is then rebinned to match the $\lambda$ sampling of the observed data. Rebinning conserves the the total flux by integrating the convolved flux over each bin in the observed wavelength grid, such that the resulting flux in each bin is given by
\begin{equation}
    F_i=\frac{1}{\Delta \lambda_i} \int_{\lambda_{i,\ \mathrm{min}}}^{\lambda_{i,\ \mathrm{max}}}F'(\lambda)d\lambda,
\end{equation}
where $F_i$ is the flux within the $i$-th rebinned pixel of the spectrum, and $F'(\lambda)$ is the convolved flux. Thus, the resulting model flux has both the spectral resolution and pixel sampling of the observed data.

Each observed spectrum is normalized by a linear fit to the continuum around \MgII. We use the flux in the rest wavelength ranges $2765-2780\ \Angstrom$ and $2810-2825\ \Angstrom$ to fit to the continuum level, but mask any pixels which deviate from the median of the continuum by $> 1\sigma$. We developed a code which samples the posterior probability density function (PPDF) of each model using an algorithm that adopts the Markov chain Monte Carlo (MCMC) method described by \citet{Foreman-Mackey_2013}. We assume that the logarithm of the likelihood function is given by $\chi^2/2$ for each model. Our code produces marginalized PPDFs of each of the parameters listed above. The probability densities are sampled over the following parameter intervals: $0 \leq C_f \leq 1$, $10 \leq \mathrm{log}N/\cmsq \leq 18$, $10\ \kms \leq b_D \leq 200\ \kms$, and $\lambda_{\mathrm{red},0}$ is allowed to vary within $\pm 400\ \kms$ of its rest-frame wavelength. We adopt the median of each distribution as the best-fit value, with uncertainties defined by the 16th and 84th percentiles, corresponding to the $1 \sigma$ interval. The velocity of the flows, hereafter referred to as $\Delta v$, is calculated from the central wavelength of the fitted Gaussian. All measured parameters are listed in Table~\ref{tab:flows}.

Given our spectral resolution, we model each flow with a single component. However, the absorption may arise from multiple clouds at different velocities \citep{Martin&Bouche2009}, which the model cannot individually resolve. This does not have an effect on the measured central velocity of the flow, but the values of the {\MgII} column density can only be presented as lower limits \citep{Huberty2024}. This limitation is increased by saturation of our absorption lines and partial covering, which would also reduce the measured column densities.

Subsequently, we estimated lower limits on the Hydrogen column densities associated with our flows. Following the method described in \citet{Rubin_2014}, we get the following lower limit for our Hydrogen column densities:

\begin{equation}
    N_{\mathrm{flow}}(\mathrm{\HI}) > N_{\mathrm{flow}}(\MgII) \frac{1}{\chi(\MgII)} \frac{1}{10^{\log {\rm Mg} / {\rm H}}} \frac{1}{10^{d({\rm Mg})}}
\end{equation}

where $\chi(\MgII)$ is the fraction of singly ionised magnesium relative to the total Mg, $\log {\rm Mg/H}$ is the Mg abundance and $d({\rm Mg})$ is the dust depletion factor for Mg. We observe some \MgI~absorption on top of our galaxies. However, this absorption has the same velocities as the ISM component of our analysis. Therefore, the {\MgI} absorption is most likely stellar or from the ISM rather than outflows or inflows. It is then reasonable to assume that $\chi(\MgII)=1$ in the case of our flows. We estimated the \MgII-dust depletion using the relationship from \citet{DeCia2016} ($d(\rm Mg) = -0.61 \times [\rm Zn / Fe] - 0.03$), and adopting the $[\rm Zn / Fe]$ reported by \citet{Wendt2021} for inflows ($[\rm Zn / Fe] = 0.5$) and outflows ($[\rm Zn / Fe] = 0.8$). This leads to $d(\rm Mg) = - 0.3$ for inflows and $d(\rm Mg) = - 0.5$ for outflows.

Magnesium abundances were estimated using two approaches. For three galaxies (J033438, J121453, and J155003, see Section~\ref{metallicity}), we directly measured oxygen abundances and converted them to magnesium abundances assuming solar abundance ratios \citep{Savage_Sembach1996}. For the remaining targets, we used the previously mentioned $[\rm Zn / Fe]$ value from \citet{Wendt2021}, converted them into zinc abundances via the \citet{DeCia2016} relationship ($[\rm Zn/Fe] = 1.37 \times [\rm Zn / Fe] - 1.73$), and then transformed the zinc abundances into magnesium abundances using the same solar abundance ratios. 

It is relevant to mention that since our {\MgII} equivalent widths are higher than 3~{\AA}, our targets are most likely damped \Lya~absorbers \citep[DLAs, ][]{Rao2006, Ellison2006, Berg2017}, where neutral hydrogen typically dominates the total hydrogen content. Thus, the derived $N(\mathrm{\HI})$ values are equivalent to $N(\mathrm{H})$.

\subsection{Mass outflow rate and mass loading factor derivations} \label{mass_outflow_rate}

We derive mass outflow rates on the down-the-barrel observations using methods similar to those from \citet{Rubin_2014}. Assuming the flows are a continuous stream coming from/going to the centre of their host galaxies at a constant speed $v$, we can calculate lower limits for the mass outflow rate as:

\begin{equation}
    \dot{M} = \alpha \mu m_{\mathrm{p}} N_{\mathrm{flow}}(\mathrm{H}) A_{\mathrm{flow}} \Delta v / D
\end{equation}

where $\mu m_{\mathrm{p}}$ is 1.4 times the mass of the proton, $N_{\mathrm{flow}}$(H) is the Hydrogen column density, $A_{\mathrm{flow}}$ is the projected surface area of the flows, $\Delta v$ is the velocity of the flows and $D$ is the extent of the flows. The constant $\alpha$ accounts for the geometry of the flow: we adopt $\alpha = 1/3$ for outflows, which are assumed to be biconical \citep{Weiner2009}, and $\alpha = 1$ for inflows, which are assumed to be filamentary \citep{Rubin2012}. We assumed that the flows reach at least the impact parameter of the quasar sightline, given that we detect absorption there with matching velocities (see Section~\ref{quasar}). Therefore, we use the impact parameter as the value for $D$.

The $A_{\mathrm{flow}}$ value is defined as $C_{f} \times S_{\rm g}$, where $C_{f}$ is the covering fraction of the flows and $S_{\rm g}$ is the surface area of the galaxy from its continuum emission in the g-band. To compute this latter value, we assumed the galaxies are disks and determined the area using~$\pi R_{\mathrm{e}}^2$. In the case of outflows, which typically have a biconical geometry, we multiply the disk area by a factor of two to account for the receding part of the flow. The values of $R_{\mathrm{e}}$ were obtained by computing a light profile of the galaxies using DECaLS archival r-band imaging and fitting a S\'ersic profile \citep{Sersic_1968} to them. 

Additionally, we quantify how efficiently the outflows remove gas from the galaxies relative to their star-formation by calculating their mass loading factor ($\eta$). This quantity is defined as:
\begin{equation}
    \eta = \dot{M} \: \mathrm{SFR}^{-1}.
\end{equation}

The calculated mass outflow and inflow rates, along with mass loading factors, are presented in Table~\ref{tab:flows}.

\begin{figure}
     \centering
     \includegraphics[width=\columnwidth]{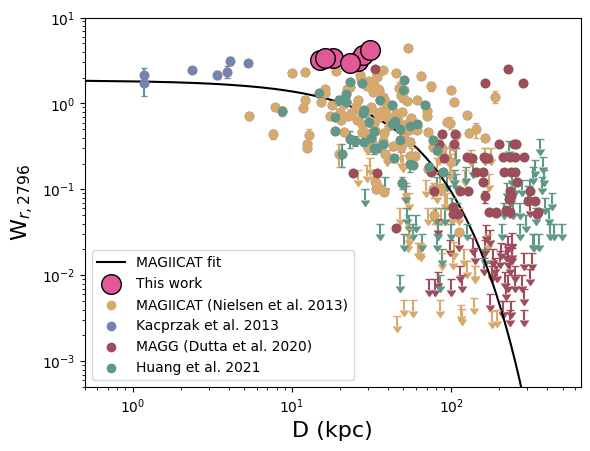}
     \caption{{\MgII} equivalent width as a function of impact parameter of our sample (large pink circles) composed of ultra-strong {\MgII} absorbers. For comparison, we also include measurements from other studies \citep{Nielsen_2013b, Kacprzak2013, Dutta2020, Huang2021}. Detections are displayed as circles, while {$3\sigma$} upper limits are shown as downward arrows. The solid black curve is the log-linear best fit to the MAGIICAT data presented in \citet{Nielsen2013a}.}
     \label{fig:ew_qso_abs}
\end{figure}
\begin{figure*}

     \begin{subfigure}[b]{0.8 \textwidth}
     \centering
     \includegraphics[width=\textwidth]{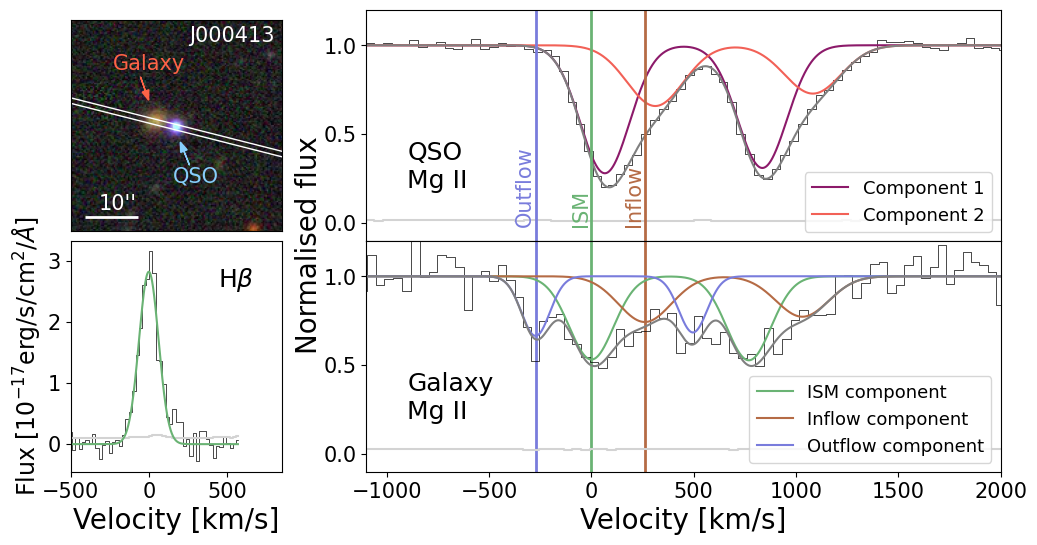}
     \end{subfigure}
     \begin{subfigure}[b]{0.8 \textwidth}
     \centering
     \includegraphics[width=\textwidth]{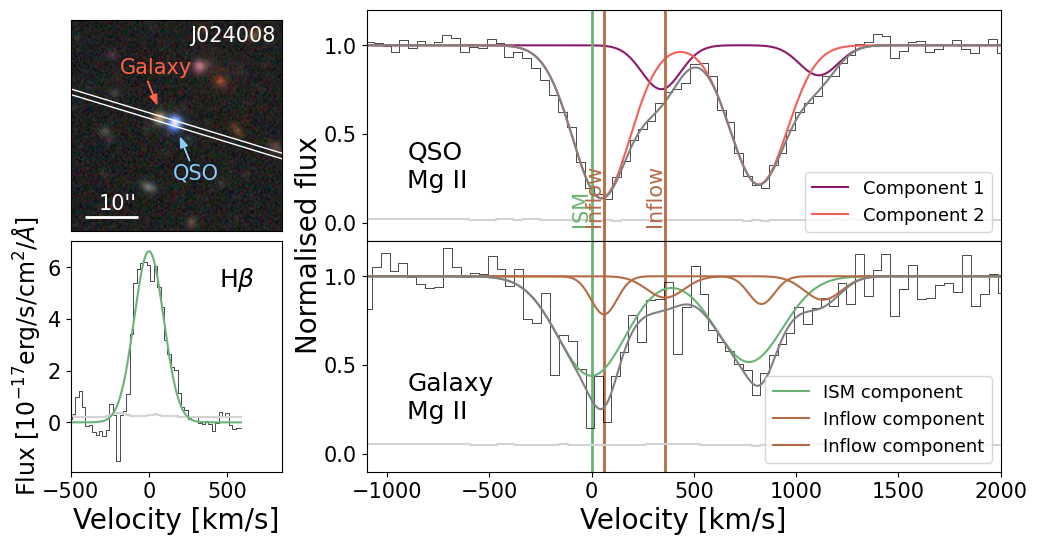}
     \end{subfigure}

   \caption{\textit{Top left panels: } DECaLS $grz$ image of each quasar field with their respective foreground galaxies. The location of the quasar and galaxy is indicated with the blue and red arrows, respectively. The white lines represent the location of the LRIS slit. \textit{Top right panels: } {\MgII}~$\lambda\lambda 2796,2803$ absorption doublet found in the spectra of the background quasars. Individual absorption components are shown in purple, orange and dark green when present. \textit{Bottom left panels: } \Hbeta~emission line of each galaxy. In the case of J155003, the {\Hbeta} emission line is not covered in our wavelength range, so we present {\Hgamma} instead. \textit{Bottom right panels: } {\MgII}~$\lambda\lambda 2796,2803$ absorption doublet of each galaxy. In all spectral panels, the zero-velocity point is defined as the central velocity of the {\Hbeta} (or {\Hgamma}) emission line. The galaxy spectra are shown in black and the 1$\sigma$ errors are presented in light grey. Individual absorption components are colour-coded: green traces the ISM, blue represents outflows, and brown indicates inflows. Vertical lines highlight the central velocity of the ISM, inflows and outflows, as labelled. The total fitted model is shown in dark grey.}
   \label{fig:mgii_outflow}
\end{figure*}

\begin{figure*}

     \ContinuedFloat
     
     \begin{subfigure}[b]{0.8 \textwidth}
     \centering
     \includegraphics[width=\textwidth]{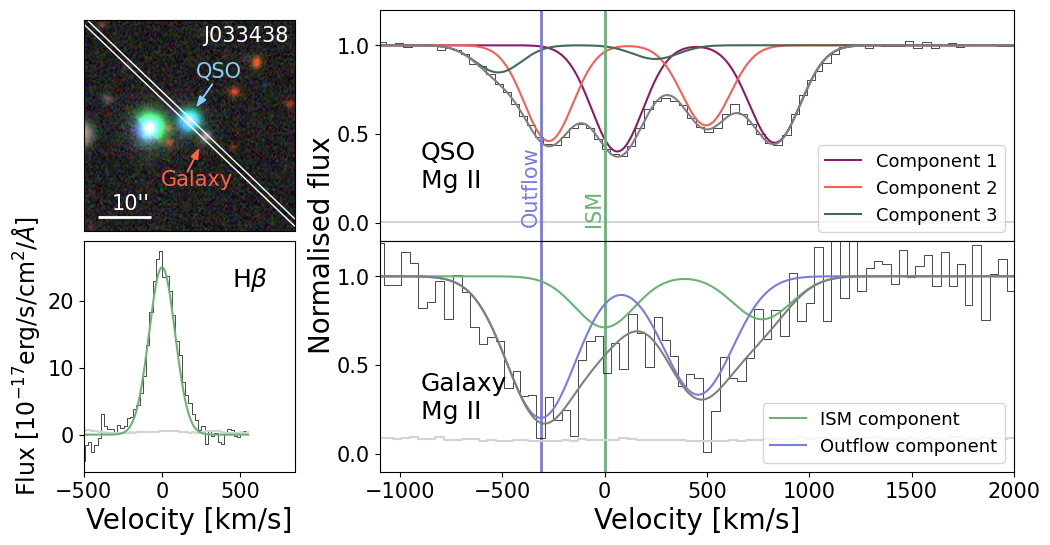}
     \end{subfigure}
     \begin{subfigure}[b]{0.8 \textwidth}
     \centering
     \includegraphics[width=\textwidth]{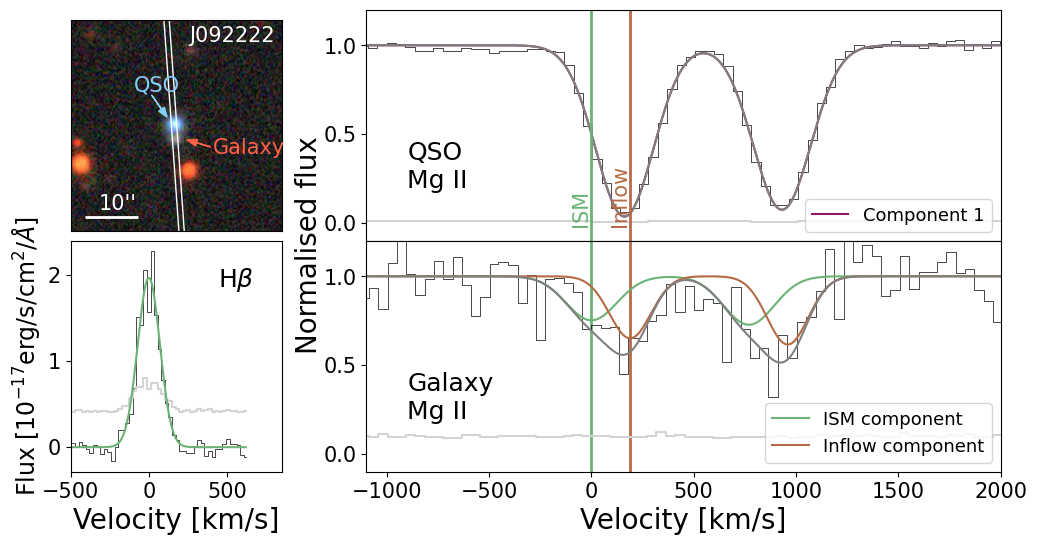}
     \end{subfigure}
     \begin{subfigure}[b]{0.8 \textwidth}
     \centering
     \includegraphics[width=\textwidth]{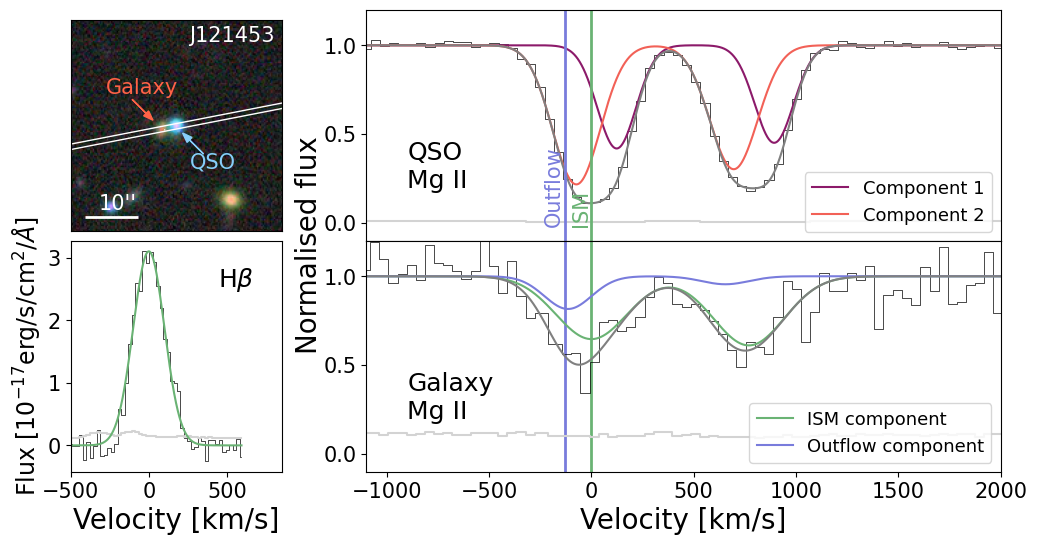}
     \end{subfigure}

   \caption{(continued)}
   \label{fig:mgii_outflow}
\end{figure*}

\begin{figure*}

     \ContinuedFloat
     \begin{subfigure}[b]{0.8 \textwidth}
     \centering
     \includegraphics[width=\textwidth]{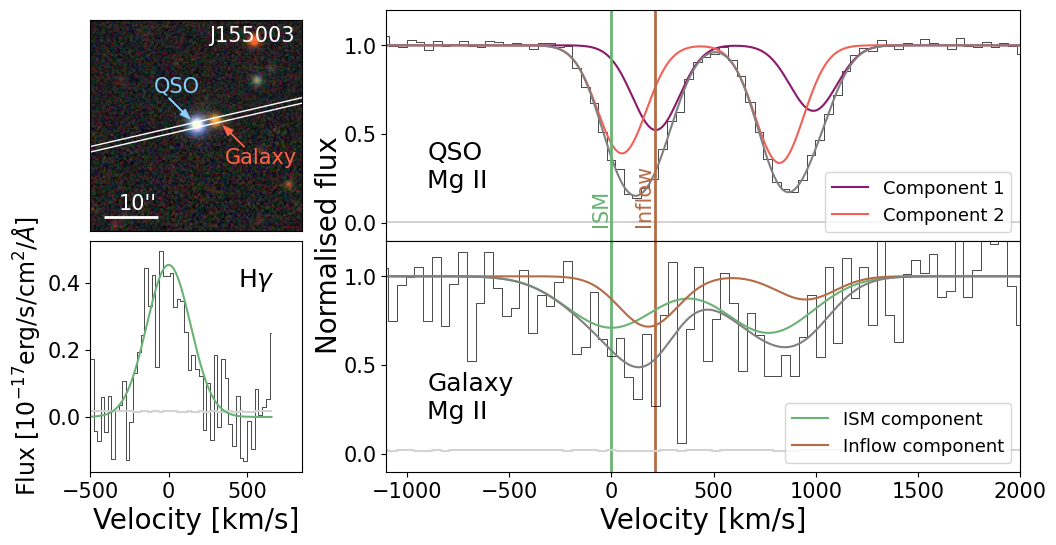}
     \end{subfigure}
     \begin{subfigure}[b]{0.8 \textwidth}
     \centering
     \includegraphics[width=\textwidth]{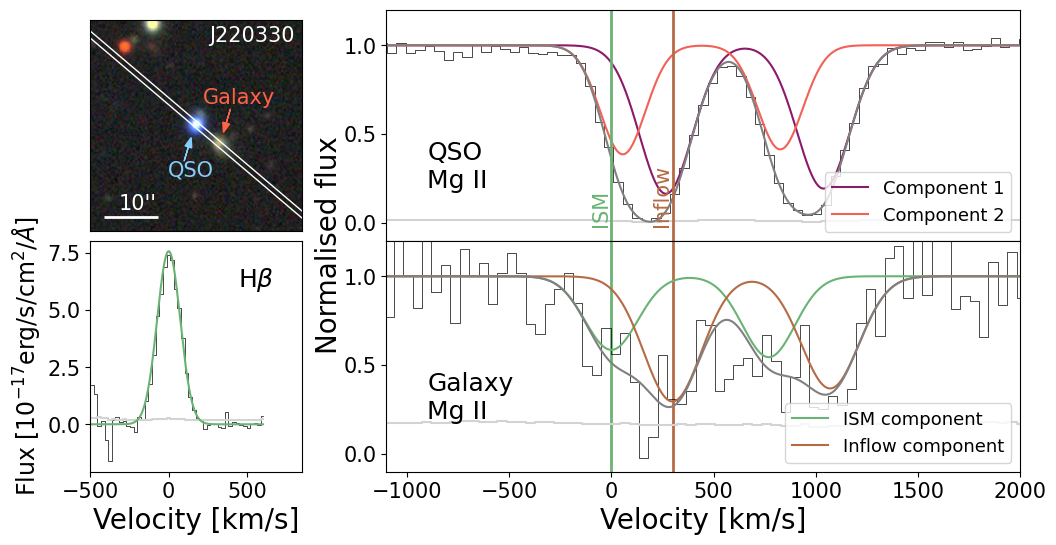}
     \end{subfigure}

   \caption{(continued)}
   \label{fig:mgii_outflow}
\end{figure*}

\section{Results} \label{results}

\subsection{General properties of the sample} \label{gen_prop}

The {\MgII} rest-frame equivalent width of our absorbers is shown as a function of impact parameter in Fig.~\ref{fig:ew_qso_abs}, along with similar measurements from literature \citep{Nielsen_2013b, Kacprzak2013, Dutta2020, Huang2021}. By design, the galaxies in our sample are associated with {\MgII} rest-frame equivalent widths higher than $3$ {\AA} detected along the nearby quasar sightlines. Ultra-strong {\MgII} absorbers have been shown to deviate from the anticorrelation of equivalent width and impact parameter \citep{Guha2022, Guha2024, Joshi2024}. To explore this possibility, we investigate the physical mechanisms that could be driving these characteristic spectral features. 

All of our galaxies have SFRs ranging from $1$ to $34$~$M_{\astrosun}$~yr$^{-1}$, with a median SFR of $2.73$~$M_{\astrosun}$~yr$^{-1}$ (see Table~\ref{tab:properties}). Given that some of these star-formation rates are high, it is interesting to address the likelihood of these galaxies having strong star formation-driven outflows. Some authors suggest that a minimum threshold of $\Sigma_{\mathrm{SFR}}$=$0.1$~$M_{\odot}$~yr$^{-1}$~kpc$^{-2}$ is required to trigger superwinds, galactic winds or strong outflows \citep{Heckman2002, Heckman2015, ReichardtChu2022}, and that $\Sigma_{\mathrm{SFR}}$ governs the amount of material expelled out from the galaxy disk. We compute $\Sigma_{\mathrm{SFR}}$ ranging between 0.004 to $0.2$~$M_{\odot}$~yr$^{-1}$~kpc$^{-2}$, with a median value of $0.01$~$M_{\odot}$~yr$^{-1}$~kpc$^{-2}$. Interestingly, only one of these galaxies (J033438) is above the expected threshold where outflows should be present. However, work by \citet{ReichardtChu2022, ReichardtChu2025} shows that outflows having velocities ranging from $100$ to $600$ {\kms} can occur for a range of $\Sigma_{\mathrm{SFR}}$, from $0.01$ to $10$~$M_{\odot}$~yr$^{-1}$~kpc$^{-2}$, an order of magnitude below the previous stated threshold. A clear way of verifying the presence of outflows is via direct down-the-barrel observations.

DECaLS images of the seven galaxy--quasar pairs, along with the position of the LRIS slit, are shown in the top left panels of Fig.~\ref{fig:mgii_outflow}. The top right panels for each galaxy display the {\MgII}~$\lambda\lambda 2796,2803$ absorption lines found along the quasar sightline. In the bottom left panels, we show {\Hbeta} emission lines associated with the ISM gas of each galaxy, while the bottom right panels present the corresponding down-the-barrel {\MgII} absorption. In Section~\ref{dtb_results} we present the results on the down-the-barrel gas flows and in Section~\ref{quasar}, we will focus in the analysis of the quasar absorption line observations.

\subsection{Down-the-barrel absorption} \label{dtb_results}

\subsubsection{Individual systems}

The systems display inflows and outflows in their down-the-barrel absorption. While Section~\ref{gen_prop} focuses on overall trends presented by these flows, here we focus on the targets individually.\\

\textit{J000413:} The host galaxy is $2.5\sigma$ below the star-forming main sequence at $z = 0.5$ \citep{Whitaker2012}, with an SFR of $2.3$~$M_{\odot}$~yr$^{-1}$ and a stellar mass of $M_{\star}=10^{10.99}$~$M_{\odot}$. As shown in the first set of panels of Fig.~\ref{fig:mgii_outflow}, this system exhibits one inflow component at $\Delta v= 261$~{\kms} and $W_{\rm r, 2796} = 0.8$~\AA, and one outflow component at $\Delta v = -272$~{\kms} and $W_{\rm r, 2796} = 0.5$~{\AA} down the barrel.\\

\textit{J024008:} The galaxy lies below the star-forming main sequence, but is still within $1\sigma$ from the relation, having an SFR of $6.6$~$M_{\odot}$~yr$^{-1}$ and a stellar mass of $M_{\star} = 10^{10.57}$~$M_{\odot}$. The second set of panels of Fig.~\ref{fig:mgii_outflow} shows that this system hosts two inflow components detected in the down-the-barrel absorption at~$\Delta v = 61$~{\kms} and $W_{\rm r, 2796} = 0.3$~{\AA}, and $\Delta v = 361$~{\kms} and $W_{\rm r, 2796} = 0.3$~{\AA}.\\

\textit{J033438:} This galaxy is $1.7\sigma$ above the star-forming main sequence, with an SFR of $33.9$~$M_{\odot}$~yr$^{-1}$ and a stellar mass of $M_{\star} = 10^{10.57}$~$M_{\odot}$. It also has an oxygen abundance of $12+\log({\rm O/H})=8.98$. Interestingly, this target has the highest SFR and $\Sigma_{\mathrm{SFR}}$ of all galaxies in this study. Moreover, our dust content estimation, necessary for the dust extinction correction, indicates that this is the most dust-rich galaxy in the whole sample, having an $A_{\mathrm{V}}=2$. As shown in the third set of panels of Fig.~\ref{fig:mgii_outflow}, this system has a strong outflow component in its down-the-barrel spectrum at $\Delta v = -311$~{\kms} and $W_{\rm r, 2796} = 3.2$~{\AA}, along with a weaker ISM component.\\

\textit{J092222:} The host galaxy is $2.2\sigma$ below the star-forming main sequence, having a SFR of $1.0$~$M_{\odot}$~yr$^{-1}$ and a stellar mass of $M_{\star} = 10^{10.26}$~$M_{\odot}$. This system, presented in the fourth set of panels of Fig.~\ref{fig:mgii_outflow}, shows a strong inflow component at $\Delta v = 189$ {\kms} and $W_{\rm r, 2796} = 0.8$~{\AA}.\\

\textit{J121453:} The galaxy lies $1.8\sigma$ below the star-forming main sequence, beyond the $1\sigma$ scatter, with an SFR of $2.7$~$M_{\odot}$~yr$^{-1}$ and a stellar mass of $M_{\star} = 10^{10.71}$~$M_{\odot}$. We also find an oxygen abundance value of $12+\log({\rm O/H})=8.98$. In the fifth set of panels of Fig.~\ref{fig:mgii_outflow}, we show that this system has an outflow component at $\Delta v = -199$~{\kms} and $W_{\rm r, 2796} = 0.5$~{\AA}.\\

\textit{J155003:} The host galaxy is $4.1\sigma$ below the star-forming main sequence, with an SFR of $1.7$~$M_{\odot}$~yr$^{-1}$ and a stellar mass of $M_{\star} = 10^{11.62}$~$M_{\odot}$. Interestingly, this is the most massive galaxy in our sample, and absorption features around the Balmer series lines suggest that this is a post-starburst galaxy \citep{Wu2018}. An oxygen abundance of $12+\log({\rm O/H})=8.76$ is measured in this galaxy. As displayed in the sixth set of panels of Fig.~\ref{fig:mgii_outflow}, this galaxy exhibits one inflow component at $\Delta v = 213$ {\kms} and $W_{\rm r, 2796} = 1.8$~{\AA}. It is important to note that in this case {\Hbeta} was contaminated by telluric features, so we used the {\Hgamma} emission line to derive SFR, $\Sigma_{\rm SFR}$, oxygen abundances, and to constrain the ISM central velocity and velocity dispersion.\\

\textit{J220330:} The host galaxy is below the star-forming main sequence, but still within its $1\sigma$ scatter, with an SFR of $3.6$~$M_{\odot}$~yr$^{-1}$ and a stellar mass of $M_{\star} = 10^{10.22}$~$M_{\odot}$. As presented in the seventh set of panels in Fig.~\ref{fig:mgii_outflow}, this system features a strong inflow at $\Delta v = 299$ {\kms} and $W_{\rm r, 2796} = 2.3$~{\AA} in the down-the-barrel spectroscopy. \\

In summary, for our seven targets, we find outflows in three of the galaxies: J000413, J033438, and J121453, while, by contrast, five galaxies have strong inflows: J000413 (which contains inflows and outflows simultaneously), J024008, J092222, J155003, and J220330 (see Fig.~\ref{fig:mgii_outflow},).

\subsubsection{Physical properties of the down-the-barrel flows} \label{phys_prop}

Results of the absorption line modelling described in Section~\ref{vp_fitting} are presented in Table~\ref{tab:flows}. The reported outflow and inflow velocities correspond to the central velocity of each Voigt profile relative to the ISM velocity, referred to as $\Delta v$. Cool phase outflow velocities typically range from $-50$ to $-800$ {\kms} \citep{Steidel2010, Rubin_2014, Du2016}, and our outflows fall within this range, with measured velocities of $\Delta v = -273$ {\kms} and $\Delta v = -311$ {\kms}. In contrast, simulations of accreting gas predict that inflows should exhibit lower velocities, ranging between $\sim$ $100$ to $200$ {\kms} \citep[][]{Goerdt_Ceverino2015, Carr_Scarlata2022}. However, only two of our detected inflows have velocities consistent with these predictions: J092222 at $\Delta v = 189$~{\kms} and the bluer component in J024008 at $\Delta v = 61$~{\kms}.

We find that outflows have column densities of ${\log \mathrm{N}({\MgII}) / {\rm cm}^{-2} \geq  13-14}$, with a mean value of ${\log \mathrm{N}({\MgII}) / {\rm cm}^{-2} \geq  14.2}$, Doppler parameters spanning between $b_{\rm D, out} = 54$ and $189$~{\kms}, and a mean Doppler parameter of  $b_{\rm D, out} = 125_{-23}^{+40}$~{\kms}. Their covering fractions range from $C_{f\rm,out}=0.3$ to $0.8$, with a mean of $0.5_{-0.12}^{+0.14}$, and their equivalent widths span from $W_{\rm r, 2796} = 0.5$ to 3.2~\AA, with a mean of $W_{\rm r, 2796} = 1.4 \pm 0.1$. In the case of inflows, we measure column densities between ${\log \mathrm{N}({\MgII}) / {\rm cm}^{-2} \geq  13-15}$, with a mean value of ${\log \mathrm{N}({\MgII}) / {\rm cm}^{-2} \geq 14.6}$, Doppler parameters spanning from $69$ to $200$ {\kms}, and a mean Doppler parameter of $b_{\rm D, in} = 108_{-12}^{+17}$. Inflows have covering fractions between $0.2$ and $0.6$, with a mean of $0.3_{-0.03}^{+0.10}$, and equivalent widths ranging from $W_{\rm r, 2796} = 0.3$ to 2.3~\AA, with a mean of $W_{\rm r, 2796} = 1.1 \pm 0.1$. Simulations have shown that outflows have higher average covering fractions \citep[0.1 to 0.9;][]{Faucher-Giguere2015} than inflows \citep[0.01 to 0.5;][]{Faucher-GiguereKeres2011, Wright2021}, and our detected flows display the same trend, although the difference is not significant.

\begin{figure}
     \centering
     \includegraphics[width=\columnwidth]{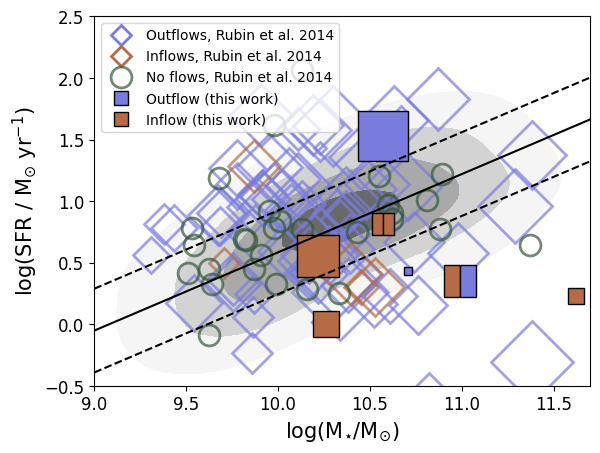}
    \caption{Location of our targets relative to the star-forming main sequence (SFMS). Grey contours represent the density distribution of star-forming galaxies from \citet{Barro2011}. The black solid line shows the SFMS at $z=0.5$ from \citet{Whitaker2012}, while the black dashed lines represent the $1\sigma$ scatter around this relation. Our targets are shown as blue and orange squares, depending on whether we detect outflows or inflows in them, respectively. The sizes of these squares are proportional to the \MgII~equivalent width of the flow in the down-the-barrel observations. In the case of J000413 and J024008, where two flow components are present, the square size is proportional to the component with the largest equivalent width. The blue and orange diamonds and the green circles correspond to the \citet{Rubin_2014} sample, colour coded in the same way as our squares, while green circles represent targets where no flows are detected. There are no such cases in the present work. These diamonds are also scaled proportionally to the \MgII~equivalent width of the flow. Their sample contains galaxies at a similar redshift to ours.}
     \label{fig:sfr_vs_m}
\end{figure}

To better understand the physical properties of the galaxies, we study the location of our galaxies relative to the star-forming main sequence at $z = 0.5$ \citep{Whitaker2012}, as shown in Fig.~\ref{fig:sfr_vs_m}. The grey contours represent the sample from \citet{Barro2011}, while our sample is displayed as squares. Interestingly, all of our inflow galaxies lie below the main sequence, suggesting that they are going through a low star-formation stage and the inflowing gas might trigger more star-formation in the future. \citet{Rubin_2014} presented outflow and inflow detections using LRIS down-the-barrel spectroscopy of 105 galaxies, having a detection rate of $66$\% for outflows and $6$\% for inflows. These galaxies have $0.3 \leq z \leq 1.4$, with a median redshift of $0.6$; their stellar masses ranges {$9.0<\log M_{\star} / M_{\odot}<11.6$}; and SFRs spanning from $0.1$ to $116$~M$_{\odot}$~yr$^{-1}$. Given the similarities between this sample, displayed in Fig.~\ref{fig:sfr_vs_m} as diamonds and green circles, and our own, it is possible to make a direct comparison between the two. Notably, when combining the inflow detections of both samples, we find that eight out of a total of ten inflows occur below the star-forming main sequence, contrary to previous results found in the literature \citep{Langan2023}, and pointing towards the idea that inflows are a mechanism for refuelling galaxies. The two exceptions that lie above the main sequence might represent an unclassified merger or a galaxy undergoing a major accretion event.

To further contextualise our inflow and outflow detections within the literature, we calculate their $|\Delta v_{\rm max}|$, a key parameter used by \citet{Rubin_2014} to describe the flow kinematics. This quantity is defined as $|\Delta v_{\rm max}| = |\Delta v| + b_{\rm D, flow} / \sqrt{2}$, where $\Delta v$ is the central velocity of the flow and $b_{\rm D, flow}$ is the Doppler parameter of the flow, which is listed in Table~\ref{tab:flows}. $|\Delta v_{\rm max}|$ provides us with a quantitative measure of the maximum velocity of the bulk of the gas.

\begin{figure}
     \centering
     \includegraphics[width=\columnwidth]{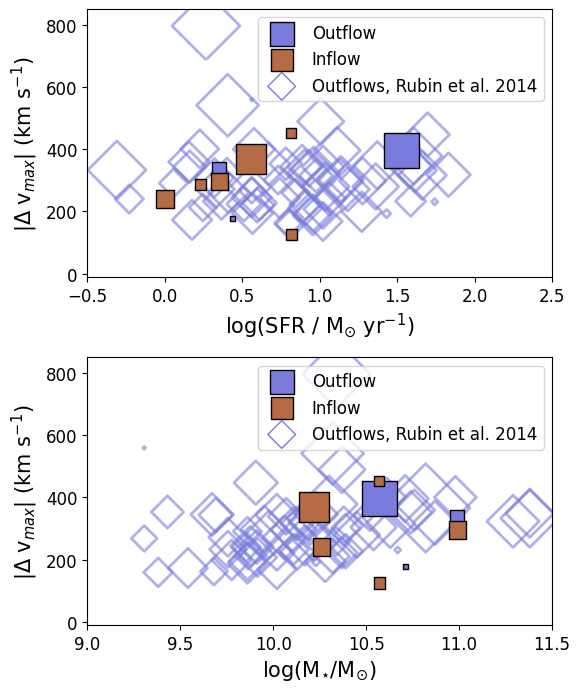}
    \caption{\textit{Top: }Outflow and inflow |$\Delta$v$_{\rm max}$| vs. star-formation rate. The inflow and outflow targets are displayed as blue and orange squares, respectively. The outflow targets from \citet{Rubin_2014} are shown as blue diamonds. The sizes of squares and diamonds are proportional to the equivalent width of the down-the-barrel {\MgII} absorption. \textit{Bottom: }Outflow and inflow |$\Delta$v$_{\rm max}$| vs. stellar mass.}
     \label{fig:vmax_vs_sfr}
\end{figure}

We analyse the relationship between $|\Delta v_{\rm max}|$ and other physical properties of the host galaxies. Fig.~\ref{fig:vmax_vs_sfr}~\textit{(top)} displays $|\Delta v_{\rm max}|$ of our targets, shown as squares, as a function of SFR, while Fig.~\ref{fig:vmax_vs_sfr}~\textit{(bottom)} shows $|\Delta v_{\rm max}|$ as a function of the host galaxies' stellar mass. The size of these squares is scaled with the equivalent width of the down-the-barrel flows. We compare these properties with those of \citet{Rubin_2014}, shown as blue diamonds in the figure. To test whether this sample is statistically different to ours, we use a 2D-Kolmogorov-Smirnov (2D-KS) test using the public code \textsc{ndtest}\footnote{Written by Zhaozhou Li, \url{https://github.com/syrte/ndtest}}. When comparing the $|\Delta v_{\rm max}|$ versus SFR, we obtain a p-value of 0.03, and for $|\Delta v_{\rm max}|$ versus stellar mass, the p value is 0.01. These results suggest that the null hypothesis that the two samples are drawn from the same population can be ruled out. This difference is likely driven by the fact that our sample exclusively contains ultra-strong {\MgII} absorbers, whereas \citep{Rubin_2014} does not consider information about equivalent widths at CGM distances.

While the central and maximum velocities of the flows are important to understand the kinematics of the systems, it is equally important to compare them to the escape velocity of their host galaxies at the virial radius. This comparison can help us determine whether the outflows have enough energy to escape their host galaxies' gravitational potential and enrich the IGM. Additionally, if the accreting gas has higher velocities than the escape velocity of its host galaxy, it would suggest that the inflow is not a corotating filament, and it is driven by other mechanisms instead, such as galaxy mergers.

\begin{figure}
     \centering
     \includegraphics[width=\columnwidth]{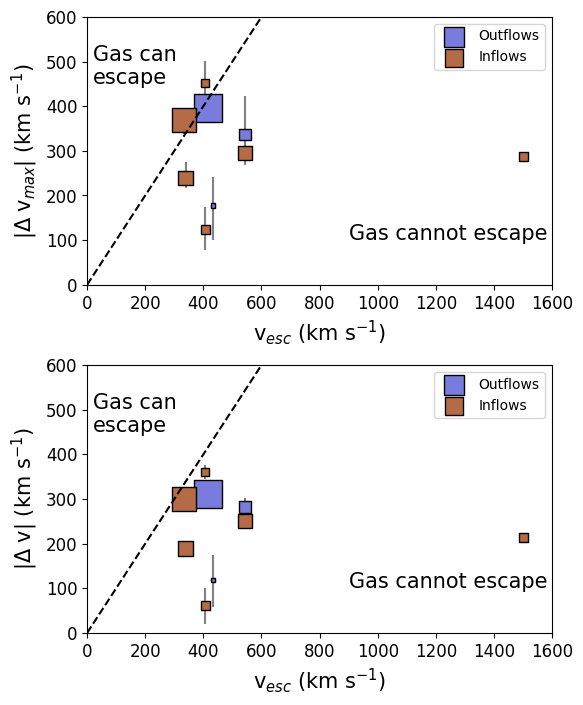}
     \caption{\textit{Top:} comparison of outflow $|\Delta v_{\rm max}|$ and $v_{\rm esc}$ of our host galaxies. \textit{Bottom:} $\Delta v$ and $v_{\rm esc}$ of the galaxies. The $v_{\rm esc}$ at the virial radius was calculated using the halo mass of the galaxies. The targets are shown as blue and orange squares, and their sizes are proportional to their {\MgII} equivalent widths. The black dashed line separates the area where the gas can escape from the area where the gas cannot do so.}
    \label{fig:vmax_vs_vesc}
\end{figure}

Fig.~\ref{fig:vmax_vs_vesc}~\textit{(top)} displays $|\Delta v_{\rm max}|$ of our targets, shown as squares, as a function of their $v_{\rm esc}$, with a dashed line separating the area where gas can potentially escape the galaxy and the area where the gas remains gravitationally bound. The escape velocity of our targets was estimated by assuming $v_{\rm esc} \simeq 3 v_{\rm circ}$ \citep{Weiner2009}, with $v_{\rm circ} = \sqrt{G M_{\rm h} / R_{h}}$. As shown in the figure, the outflow velocities of J000413, J033438 and J121453 are lower than the escape velocities of their corresponding host galaxies, indicating that this outflowing material cannot leave the CGM of their host galaxies and join the IGM. Similarly, the inflows in J000413, J024008, J092222 and J155003 display $|\Delta v_{\rm max}|$ that are lower than the escape velocity of their host galaxies, which is expected in the case of filamentary accretion. Conversely, the inflow in J220330 also has a $|\Delta v_{\rm max}|$ higher than $v_{\rm esc}$ of the galaxy, suggesting that the higher-velocity portion of this inflow does not belong to a corotating filament and is driven instead by other mechanisms.

However, $|\Delta v_{\rm max}|$ only provides information about the highest velocity gas, whereas our goal is to understand the behaviour of the bulk of the gas. In order to do that, we also compare $|\Delta v|$ with $v_{\rm esc}$, as shown in Fig.~\ref{fig:vmax_vs_vesc}~\textit{(bottom)}. In this case, the bulk of the gas in all of our targets has lower velocities than their host galaxy's escape velocity, consistent with gas that is bound to the disk and will likely continue to flow directly into it.

\subsubsection{Mass outflow and inflow rates}

To further understand the properties of these flows, we calculate mass outflow and inflow rates (see Section~\ref{mass_outflow_rate}). Results of these calculations are presented in Table~\ref{tab:flows}. Mass outflow rate lower limits range from $\dot{M} \geq -0.003$ to $-1.64$~$M_{\odot}$~yr$^{-1}$, while mass inflow rate lower limits span a range of about two orders of magnitude, from $\dot{M} \geq 0.01$ to $0.66$~$M_{\odot}$~yr$^{-1}$. However, it is important to note that these measurements have large sources of systematic errors, such as assumptions about the extent of the flows, the morphology of the gas, and the conversion between $N(\MgII)$ and $N(\HI)$ (see Section~\ref{mass_outflow_rate}). As such, these values should be interpreted as order-of-magnitude estimates.

\begin{figure}
     \centering
     \includegraphics[width=\columnwidth]{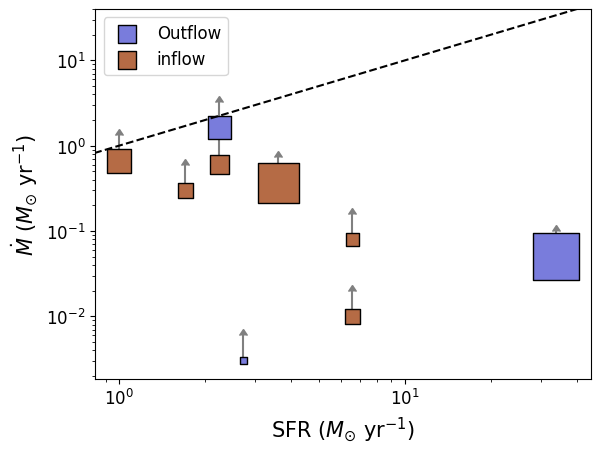}
     \caption{Down-the-barrel mass outflow/inflow rate as a function of SFR. The size of the squares is proportional to the {\MgII} equivalent width of the flow. The black dashed line represents the 1:1 ratio line. Targets sitting on this line are in a state of equilibrium, where either the inflowing gas replenishes the material used to form stars, or the outflowing gas depletes galaxies of their gas.}
    \label{fig:dm_dt_vs_sfr}
\end{figure}

Studying the mass loading factors of these flows shows us the balance, or imbalance, that the flows produce in the star-formation of the galaxies. These values are visually represented in Fig.~\ref{fig:dm_dt_vs_sfr}. All outflow targets have mass loading factors below unity, and when combined with the fact that $\Delta v \leq v_{\rm esc}$, this suggests that the galaxies retain most of their gas. Moreover, all the inflow galaxies display mass loading factors below unity, indicating they will not be able to maintain their current level of star-formation indefinitely. For these galaxies, we estimate gas depletion timescales ranging from $\eta = 0.6$ to $5$~Gyr, which are consistent with previous measurements for star-forming galaxies \citep{Saintonge2017}.

Previous studies have found that mass loading factors tend to decrease with increasing stellar mass \citep{Chisholm2017, Hamel-Bravo2024}. However, among our three outflow targets, we do not observe such a trend. Although this contrasts with previous findings, we caution that our sample size is small and larger samples are necessary to test this relationship.

\subsection{Background quasar absorption} \label{quasar}

The {\MgII} absorption lines in the background quasars at impact parameters ranging between $D=15$ to $31$~kpc are shown in Fig.~\ref{fig:mgii_outflow}~\textit{(top right)}. When analysing the {\MgII} absorption along the quasar sightline, we find that two velocity components are sufficient to explain the absorption in five of the targets: J000413, J024008, J121453, J155003, and J220330. However, J092222 and J033438 need one and three velocity components, respectively.

\begin{figure}
     \centering
     \includegraphics[width=\columnwidth]{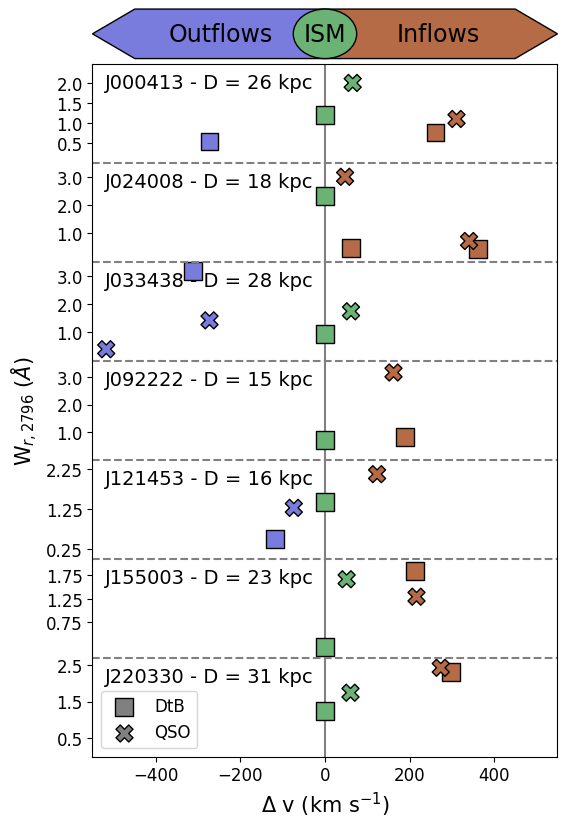}
     \caption{Comparison between different velocity components found in the {\MgII} absorption. Components found in the QSO absorption are displayed as crosses, while the components measured down-the-barrel are shown as squares. The impact parameters of the quasar are labelled on each panel. Markers coloured in blue, brown and green are associated with outflows, inflows and ISM, respectively. The zero-velocity line, shown as a black solid line, is measured from the ISM velocity of the host galaxy. The y-axis represents the rest-frame equivalent width of each velocity component.}
     \label{fig:vel_comp}
\end{figure}

Fig.~\ref{fig:vel_comp} presents a velocity comparison between {\MgII} absorption features down-the-barrel, shown as squares, and those along the QSO sightlines, displayed as crosses. It is interesting to note that almost all down-the-barrel flows, except the outflow in J000413, seem to have a counterpart within $v = 65$~{\kms} in the QSO absorption. In the case of the down-the-barrel outflow in J033438, its quasar counterpart has a velocity offset of $v = 38$ {\kms} with respect to the outflow. Moreover, there is an additional blueshifted component in the QSO absorption, which has a velocity offset of $v = -518$~{\kms} with respect to the systemic velocity and $v = -206$ {\kms} with respect to the outflow. For J000413, J092222 and J220330, the inflow velocities have relative velocity offsets of $v = 50$, $-28$ and $-25$ {\kms}, respectively. Additionally, both down-the-barrel inflow components in J024008 have counterparts in the QSO absorption, with a velocity offset of $v = 13$ and $20$ {\kms} for the lower velocity and higher velocity inflows, respectively. This correspondence suggests that the kinematic structures observed in the galaxies extend out to large distances ($D = 15 - 31$~kpc), supporting the use of ultra-strong {\MgII} absorbers as tracers of galaxy kinematics.

Additionally, we examine the equivalent widths of these velocity components in Fig~\ref{fig:vel_comp}. For inflows, the equivalent widths of the down-the-barrel and quasar components tend to match closely: their offsets are $W_{\rm r, 2796} = 0.3$, $-0.5$, 0.1, and 0.3~{\AA} for J000413, J155003, J220330, and the redder component on J024008, respectively.  This result can be interpreted as a coherent inflowing structure extending from the galaxy to the quasar sightline. However the other two inflows do not have this trend, having equivalent widths offsets of $W_{\rm r, 2796} = 2.4$ and 2.3~{\AA} for J092222 and the bluer component of J024008. It is also interesting to note that quasar sightlines tend to have larger inflow equivalent widths that their down-the-barrel counterparts, suggesting that gas is being lost on its way to the galaxy. Similarly, the ISM component tends to show stronger absorption in the quasar sightline, having equivalent width offsets of $W_{\rm r, 2796} = 0.8$, 0.8, 1.4, and 0.5~{\AA} for J000413, J033438, J155003 and J220330, respectively. For the outflows, only two of them have counterparts. The two counterpart pairs have equivalent width offsets of $W_{\rm r, 2796} = -1.7$ and 0.8~{\AA} for J033438 and J121453, respectively. The lack of a consistent trend suggests that outflows may be more varied in their properties down-the-barrel compared to quasar sightline distances. Moreover, this result suggests that outflows have a more fragmented or clumpy structure.

\section{Discussion} \label{discussion}

We analysed a sample of seven ultra-strong {\MgII} absorbers associated with isolated galaxies. These systems are thought to be associated with star formation-driven outflows \citep{Guha2022}. However, our analysis of the down-the-barrel gas kinematics revealed that only three of the host galaxies are currently exhibiting outflows, while five show clear signatures of significant inflow activity. This was an unexpected result, given our sensitivity to detect {\MgII} absorption features with equivalent widths as low as $0.1$~\AA. Previous studies, such as \citet{Rubin_2014} have shown that {\MgII} outflows typically have equivalent widths greater than $0.5$~\AA, well above our detection threshold. Therefore, if strong outflows were present in the remaining galaxies, we would expect to detect them. In the following subsections, we investigate the potential drivers of this unexpected inflow behaviour and consider possible origins for the ultra-strong absorption features. 

\subsection{Causes of the ultra-strong absorption} \label{causes_usmgii}

Two hypotheses have been proposed to explain the existence of ultra-strong absorbers: first is that they originate from interactions between galaxies \citep{Rubin2010, Gauthier2013, Nielsen2022}, and second is that they originate from star formation-driven outflows \citep{Rubin2010, Nestor2011, Nielsen2022, Joshi2024}. Although all of our galaxies have been identified as isolated, only three of them are currently driving outflows. This leads us to the question of what is primarily causing this ultra-strong absorption.

One possible explanation is that they are not currently undergoing outflows, but they did in the past. Outflows have typical velocities that range between $-50$ {\kms} and $-800$~{\kms} \citep{Rubin_2014, Du2016}. Assuming outflows travel at a constant velocity at the lower end of the expected range, the time it would take an outflow to travel from the galaxy centre to the impact parameter of the quasar (15 kpc $\leq D \leq$ 31 kpc) would be $0.4$, $0.3$, $0.4$ and $0.6$~Gyr for J024008, J092222, J155003, and J220330, respectively. The typical timescale of a starburst episode is $0.2-0.4$~Gyr \citep{McQuinn2009}, so it is possible that these galaxies underwent a starburst episode which has already finished and, as a result, there is no active outflow anymore.

An alternative explanation is the presence of an undetected companion in the vicinity of the galaxies. \citet{Guha2022} identified these galaxies as isolated by systematically looking for sources at the redshift of the absorption using DECaLS imaging, which have a magnitude limit of $m_{r} < 23.4$ \citep{Dey2019}, meaning that an undetected companion would have a magnitude equal or fainter than this value. In order to calculate the maximum mass this possible undetected companion could have, we use the mass--luminosity relation from \citet{Bell2003}:

\begin{equation}
    \log M_{\star}/L_{r} = -0.306 + 1.097 (g - r),
\end{equation}

where $M_{\star}$ is the galaxy's stellar mass and $L_{r}$ is its $r$-band luminosity. We assumed a $(g - r)$ colour of $0.5$, as it is an average value for a star-forming galaxy \citep[e.g., ][]{Taylor2015}. This calculation led to a maximum undetected companion stellar mass of {$\log (M_{\star} / M_{\odot})=9.8$, $9.7$, $9.5$, $9.8$ and $9.5$} for the J000413, J024008, J092222, J155003 and J220330 fields, respectively. Minor mergers, defined as having mass ratios of 1:4 or lower, require at least $3$~Gyr for a satellite to lose all of its specific angular momentum and fully merge with its host galaxy \citep{Boylan-Kolchin2008}. Our estimated companion masses are consistent with this definition, having mass ratios of 1:15, 1:7, 1:5, 1:67 and 1:5 for J000413, J024008, J092222, J155003 and J220330, respectively. At the current mass inflow rates observed in our sample, the accretion of these hypothetical companions would take $5$ to $33$~Gyr, a timescale consistent with minor mergers.

However, $\Delta v$ of all of our inflows are lower than the escape velocities of their host galaxies, suggesting these inflows are most likely corotating filaments or recycled accretion, rather than mergers. Simulations have found that gas accreted through mergers exhibits higher inflow velocities, often exceeding 200~{\kms} \citep{Sparre2022}, whereas filamentary accretion typically occurs between 100 and 200~{\kms} \citep{Goerdt_Ceverino2015, Carr_Scarlata2022}. Nonetheless, some nuance is required when interpreting these results, given that the observed line-of-sight velocity depends on the viewing angle of the system. It is therefore necessary to study the systems individually to understand the origin of the accreted gas.

\subsection{Prevalence of inflows in the sample}

While outflows appear to be ubiquitous in star-forming galaxies \citep{Weiner2009, Rubin_2014, Zhu2015}, inflows seem to be more elusive. This is partly due to inflows having lower velocities, typically ranging from $100$ to $200$ {\kms} \citep{Goerdt_Ceverino2015, Carr_Scarlata2022}, compared to star formation-driven outflows traced by cool gas phase, which commonly range between $-50$ and $-800$ {\kms} \citep{Rubin_2014, Du2016}. The lower velocity of inflows makes it difficult to disentangle them from ISM absorption in low-spectral resolution observations. Furthermore, there is both observational \citep{Martin2019, Banerjee2025} and simulation \citep{Dekel2009, Kimm2011, Fumagalli2011} evidence that inflows have a filamentary structure, which leads to lower covering fractions. Simulations suggest that these inflows are limited to covering fraction values between $0.01$ and $0.4$ \citep{Faucher-GiguereKeres2011, VandeVoort2019}, further complicating their detection. Moreover, evidence suggests that inflows tend to occur along the major axis of galaxies, typically at azimuthal angles below $40^{\circ}$ \citep{Zabl2019}. This geometric preference poses an observational limitation for their detection \citep{Rubin2012, Weldon2023}. It is due to these reasons that we explore different explanations for our discovery of five inflows in a sample of seven galaxies.

One possible explanation for this finding is that galaxies associated with ultra-strong absorption tend to be located in denser environments \citep{Joshi2024}. This could potentially attract filaments from the IGM to be accreted by the galaxy disk. This once again highlights the importance of studying the metal content of these inflows, in order to understand whether they have a cosmic origin or if they are recycled gas from the galaxy itself. If the metallicity of the inflowing gas is comparable to that of the host galaxy, it is likely to be recycled gas, whereas significantly lower metallicities would suggest a cosmic origin. In addition to metallicity, understanding the geometry of the flows relative to the host galaxy is also crucial, given that inflows occur preferentially along the major axis of the disk \citep{Zabl2019}. However, it is important to highlight that the available imaging of our galaxies is not sufficiently deep to constrain their inclinations. Future observations will be necessary to determine whether our detected inflows lie along the major axis of their host galaxies, and therefore follow the trend reported in the literature.

We identify inflows with $W_{r, 2796} \geq 0.3$~\AA~in five of our seven galaxies. This corresponds to a detection rate of 71\%, which is remarkably high when compared to previous studies such as \citet{Rubin_2014}, who reported a detection rate of only 6\% on a similar sample. The probability of detecting inflows in five or more galaxies out of seven, assuming a true detection rate of 6\%, is just 0.002\%, making our result highly unlikely and particularly unexpected. Our unexpectedly high inflow detection rate motivates a closer examination of each system to better understand the causes of these gas flows. In the following subsection, we analyse each inflow galaxy individually, in order to shed light on their origins.

\subsection{Origin of detected inflows}

Each galaxy in our sample presents unique kinematic characteristics, such as differences in inflow velocities, column densities, covering fractions, and mass inflow rates, suggesting a variety of mechanisms being responsible for this observed accretion. We analyse the properties of these inflows and their host galaxies to investigate the origins of the inflowing gas.  \\

\textit{J000413:} we find evidence for a galactic fountain in this target. Both inflows and outflows are observed down-the-barrel, with $|\Delta v_{\rm max}|=337$ and $261$~{\kms}, respectively, below the galaxy escape velocity of $v_{\rm esc} = 543$~{\kms}, indicating that the gas is gravitationally bound. In such a scenario, inflowing and outflowing gas would have metallicities and Fe abundances similar to the host galaxy, unlike cosmic accretion, which would show lower values. The inflow also has a counterpart in the quasar sightline, closely matched in velocity ($v = 50$~\kms) and equivalent width ($W_{\rm r, 2796} = 0.3$~\AA), suggesting a coherent structure from the galaxy to the sightline. In contrast, the outflow lacks a clear quasar counterpart, implying a clumpier structure. Additional data is needed to confirm this hypothesis. \\

\textit{J024008:} this galaxy features two inflow components, at $\Delta v = 61$ {\kms} and $361$ {\kms}. The lower-velocity component falls within the typical range for inflows \citep[typically $< 100-200$~{\kms}, ][]{Goerdt_Ceverino2015, Carr_Scarlata2022}, likely tracing gas corotating with the galaxy disk. The higher velocity component exceeds the typical inflow velocities and the host's escape velocity of $v_{\rm esc} = 406$~{\kms}. Simulations indicate that filamentary accretion generally occurs at $\leq 200$ {\kms} \citep{VanddeVoort2012}, suggesting this gas likely originates from other mechanisms, such as minor mergers. Given the field's magnitude limit, any undetected companions would have at least one order of magnitude lower mass than the host (mass ratio $\leq 1:7$, see Section~\ref{causes_usmgii}). Notably, while both inflow components have velocity counterparts in the quasar sightline, only the redder component matches in equivalent width. \\

\textit{J092222:} the inflow has $|\Delta v_{\rm max}|=240$~{\kms}, below the galaxy's escape velocity ($v_{\rm esc} = 339$~{\kms}), suggesting the redshifted down-the-barrel absorption traces gas accreting and corotating with the disk. The mass inflow rate lower limit is $\dot{M} \geq 0.66$~$M\odot$~yr$^{-1}$, compared to the SFR of $1$~$M\odot$~yr$^{-1}$, indicating that the galaxy is no longer sustaining significant star-formation. This inflow also has a velocity counterpart in the quasar sightline, though with a large difference in equivalent width ($W_{\rm r, 2796} = 2.4$~\AA), suggesting a clumpy structure. \\

\textit{J155003:} the redshifted down-the-barrel absorption is consistent with filamentary accretion, as the inflow's $|\Delta v_{\rm max}|=287$~{\kms}, well below the host galaxy's escape velocity ($v_{\rm esc} = 1500$~{\kms}). The inflow has a quasar sightline counterpart, with small differences in velocity ($v = 25$~\kms) and equivalent width ($W_{\rm r, 2796} = -0.5$~\AA), suggesting a coherent filamentary structure. However, the available DECaLS imaging is not deep enough to determine the galaxy's inclination and orientation relative to the quasar, so further data is necessary to confirm this scenario. \\

\textit{J220330:} the inflow associated with this galaxy has $|\Delta v_{\rm max}|=369$~{\kms}, exceeding the host's escape velocity ($v_{\rm esc} = 332$~{\kms}), suggesting it is not filamentary accretion but may originate from a minor merger. Based on our analysis of possible undetected companions (see Section~\ref{causes_usmgii}), the mass ratio would likely be $1:5$ or lower. This interpretation is supported by the inflow's large covering fraction ($C_{f\rm , in} = 0.6$), unusually high compared to simulations predicting $C_{f\rm , in} = 0.01$ and $0.4$ \citep{Faucher-GiguereKeres2011, VandeVoort2019}, and by its large equivalent width, the highest in the sample. The inflow also has a quasar sightline counterpart closely matched in both velocity ($v = 25$~\kms) and equivalent width ($W_{\rm r, 2796} = 0.1$~\AA), suggesting a coherent structure. \\

Altogether, we find evidence for one galactic fountain, two cases consistent with filamentary accretion, and two likely driven by minor mergers. These results suggest that ultra-strong {\MgII} absorbers might be used to identify strong accretion events. However, the small sample size of our study highlights the need for further observations of ultra-strong {\MgII} absorber hosts and their metal content to confirm these findings. In particular, deeper imaging of the galaxies is necessary to resolve the spatial distribution of the flows, and to search for possible companions that have gone undetected in the current data.

\section{Summary and conclusions} \label{conclusions}

In this paper, we study a sample of seven galaxy--QSO pairs associated with ultra-strong {\MgII} absorbers. The galaxies are star-forming, have redshifts of $0.4<z<0.6$, have similar stellar masses of $10.2 < \log M_{\star} / M_{\odot} < 11.6$, and are identified to be isolated by \citet{Guha2022}. We observed these targets using Keck/LRIS longslit spectroscopy and find the following:

\begin{itemize}
    \item The galaxies have star-formation rates ranging from $1$ to $34$~$M_{\astrosun}$~yr$^{-1}$. However, only one of these galaxies features a $\Sigma_{\mathrm{SFR}}$ above the wind-driving threshold and is above the star forming main sequence. 
    \item Analysing down-the-barrel {\MgII} absorption, we detect outflows in three of our targets, displaying velocities spanning from  $-311$ and $-119$ {\kms}, consistent with typical outflow velocities. As expected, one of these outflows is detected in the galaxy with the highest $\Sigma_{\mathrm{SFR}}$. On the other hand, we detect inflows in five of our targets, spanning velocities from 61 to $361$ {\kms}. Interestingly, one of our targets exhibits both inflows and outflows at the same time, suggesting it features a galactic fountain. 
    \item Outflows have column densities of ${\log \mathrm{N}({\MgII}) / {\rm cm}^{-2}\geq14}$ along with covering fractions ranging from $0.3$ to $0.8$, yielding a mean of 0.5. Inflows have column densities of  ${\log \mathrm{N}({\MgII}) / {\rm cm}^{-2}\geq13-15}$, along with covering fractions ranging from $0.2$ to $0.6$, with a mean covering fraction of 0.3. Covering fractions of outflows are generally larger than those associated with inflows.
    \item Mass outflow rate lower limits range from $1.64$ to $0.003$~$M_{\astrosun}$~yr$^{-1}$, while mass inflow rate lower limits span from $0.01$ to $0.66$~$M_{\astrosun}$~yr$^{-1}$. When studying the mass loading factors of the outflows, we find that all are below unity, suggesting that these galaxies retain most of their gas. Similarly, all inflow systems have mass loading factors below unity. It is important to note, however, that the present study focuses only on the $10^4$~K photoionised gas phase. Other gas phases might also be present and accreting, potentially helping to sustain the current level of star-formation of our galaxies. 
    \item When comparing the {\MgII} absorption observed in the quasar line-of-sight with that detected down-the-barrel, we find that almost all inflows and 50\% of outflows have counterparts in the quasar line-of-sight, suggesting that ultra-strong {\MgII} absorbers are good tracers of galactic gas flows and the baryon cycle at impact parameters as high as $31$~kpc. 
    \item  We compared the equivalent width of inflow, outflow and ISM components along the down-the-barrel and quasar sightlines. Inflows generally show similar equivalent widths in both directions, consistent with coherent inflows extending to the quasar impact parameter, although two exceptions suggest a more complex structure. ISM absorption also tends to be stronger along the quasar sightline. In contrast, outflows display no consistent equivalent width trends between the two sightlines, implying a more fragmented or clumpy structure. 
    \item Outflows have been suggested as a possible cause for ultra-strong {\MgII} absorption in isolated galaxies. However, we only detect outflows in three of our seven targets. It is possible that the other galaxies underwent outflow events $0.2$ to $0.6$~Gyr in the past and we are detecting traces of these outflows in the quasar line-of-sight. Another possibility is the presence of undetected companions near our galaxies. We find that the properties of our inflows are consistent with minor mergers (1:4 mass ratio or lower). However, given that the kinematic analysis of the flows shows consistency with corotating filaments, it is necessary to analyse the systems individually to determine their origin.
    \item Detections of inflows using down-the-barrel {\MgII} absorption lines of star-forming galaxies are rare. There have only been 15 of those detections in the literature \citep{Rubin2012, Martin2012}. It is important to note that those studies focused on typical {\MgII}, rather than ultra-strong {\MgII} absorbers. Unexpectedly, we detect five inflows in a sample of seven galaxies. Although our targets appear to be isolated, it is possible that they host undetected low-mass companions with stellar masses $\leq 10^{9.5}$~$M_{\odot}$, which could contribute to the observed inflow activity. 
\end{itemize}

In conclusion, we found five inflows with similar characteristics, such as velocities and covering fractions, to those found in the literature. Given the elusive nature of inflows, this might be an indication that observing galaxies associated with ultra-strong absorbers could be used as a technique to find inflows. However, more data, such as deeper imaging of the galaxies or observations of metallicity tracers, are necessary to confirm this claim.

\section*{Acknowledgements}
We thank our referee for their helpful comments that have improved the paper.

Some of the data presented herein were obtained at the W. M. Keck Observatory, which is operated as a scientific partnership among the California Institute of Technology, the University of California, and the National Aeronautics and Space Administration. The Observatory was made possible by the generous financial support of the W. M. Keck Foundation. Observations were supported by Swinburne Keck program 2023B\_W431 and 2025A\_W368. The authors wish to recognize and acknowledge the very significant cultural role and reverence that the summit of Maunakea has always had within the indigenous Hawaiian community. We are most fortunate to have the opportunity to conduct observations from this mountain.

Parts of this research were supported by the Australian Research Council Centre of Excellence for All Sky Astrophysics in 3 Dimensions (ASTRO 3D), through project number CE170100013.

\section*{Data Availability}

The data underlying this paper will be shared following mutually agreeable arrangements with the corresponding authors.
 



\bibliographystyle{mnras}
\bibliography{refs}








\bsp	
\label{lastpage}
\end{document}